\documentclass[aip, amsmath,amssymb,preprint,graphicx]{revtex4-1}

\draft 

\usepackage{graphicx}
\usepackage{dcolumn}
\usepackage{bm}

\usepackage[utf8]{inputenc}
\usepackage[T1]{fontenc}
\usepackage{mathptmx}

\begin{document}
	
	\title{The role of standing wave in the generation of hot electrons by femtosecond laser beams incident on dense ionized target} 
	
	\author{R. Babjak}
	\email[]{babjak@ipp.cas.cz}
	\affiliation{Faculty of Nuclear Sciences and Physical Engineering, Czech Technical University in Prague, Brehova 7, 115 19 Prague 1, Czechia}
	\affiliation{Institute of Plasma Physics, Czech Academy of Sciences, Za Slovankou 1782/3, 182 00 Praha 8, Czechia}
	
	\author{J. Psikal}
	\affiliation{Faculty of Nuclear Sciences and Physical Engineering, Czech Technical University in Prague, Brehova 7, 115 19 Prague 1, Czechia}

	\date{\today}
	
	\begin{abstract}
		We demonstrate the differences in hot electron absorption mechanisms dominant in the interaction of femtosecond laser pulse with intensities $10^{18}~\rm{W/cm^2}$ and $10^{21}~\rm{W/cm^2}$ on fully ionized target with steep density profile and preplasma with moderate scale length (3 $\rm{\mu}$m). We show that acceleration of each electron starts at the moment when magnetic component of standing electromagnetic wave changes its polarity in regime without preplasma. In the presence of preplasma the stochastic heating is the dominant absorption mechanism along with the longitudinal electric field. It is observed, that wave's energy is absorbed only if standing wave is already created at the position of electron during the interaction with the pulse with intensity $10^{18}~\rm{W/cm^2}$. In the case with intensity $10^{21}~\rm{W/cm^2}$, part of electrons is pre-accelerated in front of the target before the reflection and following stochastic heating. The presence of preplasma results in electron temperatures close to or even exceeding ponderomotive scaling. At higher intensity, the re-injection of electrons previously repelled by incident wave's ponderomotive force into high-field regions is allowed if standing wave is created. 
	\end{abstract}
	
	
	\maketitle 

	\section{Introduction}
	The interaction of laser pulses with intensities exceeding $10^{18}~\rm{W/cm^2}$ with overdense thin ionized targets found many applications in recent years, e.g. laser-driven particle acceleration \cite{wilks_tnsa}, high harmonics generation \cite{norreys_hhg}, fast ignition in inertial confinement fusion \cite{craxton_icf} or laboratory astrophysics \cite{bulanov_astro}. When electromagnetic (EM) field at such intensities interacts with the plasma, fraction of the radiation is absorbed via non-linear processes present during the interaction and the rest is reflected from the overcritical electron-plasma boundary (EPB). Therefore, the understanding of underlying radiation-matter coupling mechanisms is essential for this research.
	
	Energy from the EM field that is absorbed is mostly converted into the population of so-called hot electrons which are accelerated/generated near the target surface or in the underdense preplasma formed in front of the target. Many theoretical and empirical models were proposed to explain such acceleration \cite{macchi_review,beg_scaling,kluge_scaling,wilks_abs,denavit_abs,chen_electrons,culfa_electrons}, but they differ in prediction of electron properties such as number of accelerated particles, quasi-temperature, cut-off energy or angular distribution depending on simulation/experimental parameters. The acceleration mechanisms of hot electrons seem to be the most sensitive to laser intensity, preplasma profile and angle of incidence of laser pulse on the target. 
	
When steep density profile is present, the acceleration is strongly influenced by the quasi-static electric field created at the EPB \cite{debayle,bulanov_foil} and by the electromagnetic field in the form of standing wave created due to the interference of incident and reflected laser pulse \cite{kemp,2a0,huller_electrons}. 
When preplasma \cite{esirkepov_preplasma,preplasma_had} is present, the coupling is explained mostly by stochastic heating \cite{stoch_paradkar,stoch_sheng,sentoku_stoch,chopineau_stoch} or Direct laser acceleration (DLA) mechanism \cite{dla,lida}. 
	
This work aims to provide further insight into the regime of interaction when femtosecond laser pulse is incident normally on the target. 
In the case of steep plasma density profile, we discuss in detail the ejection of electrons into the vacuum caused by the electrostatic field created at the vicinity of EPB and their acceleration in the standing wave created in front of the target.
When intensity of incident laser pulse is higher, we demonstrate the effect of target surface deformation due to strong ponderomotive pressure. 
In preplasma, hot electron acceleration mechanism strongly differs compared to the acceleration in the vicinity of EPB at sharp interface and leads to electron temperatures close to or even exceeding ponderomotive scaling. 
We clearly demonstrate the necessity of standing wave for moderate length preplasma (in order of $\mu$m ) in order to accelerate electrons and the presence of stochastic heating and describe why longer preplasma results in higher electron energies. 
For higher intensity, we show that in the presence of standing wave, previously repelled electrons by the ponderomotive force can be re-injected into the high-field regions. We also propose an explanation of higher temperatures of hot electrons in the case od p-polarization compared to s-polarization in the target with preplasma.
	
	\section{Methods and simulations parameters}
	In order to study hot electron generation, the 2D version of Particle-In-Cell code Smilei \cite{smilei} was used. Linearly p-polarized laser pulse of wavelength $\lambda=800$ nm with intensities $10^{18}~\rm{W/cm^2}$ and $10^{21}~\rm{W/cm^2}$ at maximum (i.e., with dimensionless pulse amplitudes $a_0 \approx 0.68$ and $a_0 \approx 21.6$, respectively) was incident on target with sharp density profile and on the target with preplasma in front of the dense thin foil. The combination of density profiles and pulse intensities resulted in 4 investigated scenarios. For simplicity, plasma containing only electrons and protons was assumed. The density of overdense foil was $\approx30~n_c$ (which corresponds to fully ionized hydrogen in the density of solid state), where $n_c$ is critical density. Scale length of exponentially increasing density of preplasma $n=n_c\times\exp(x/L)$ for $-14.5~\rm{\mu m}<x \leq 0 ~\rm{\mu m}$ was $L=3\rm{\mu}$m, until reached the overdense part of the target with constant maximum density located at $0~ \rm{\mu m}<x<5~ \rm{\mu}$m. Target with steep density gradient contained only overdense thin foil and the above defined preplasma was put in front of the target for another studied cases. The size of grid cell was set to $16\times16~\rm{nm}$, 50 macroparticles per cell were used for overdense part and preplasma contained 16 macroparticles per cell due to its lower density. Laser pulse was propagating towards the target in x direction from the boundary located at $x_{min}$ ($x_{min}=-12~\rm{\mu}$m for target without preplasma and $x_{min}=-15~\rm{\mu}$m for target with preplasma) with the beam axis at the position $y=0~\rm{\mu}$m, it had a trapezoidal temporal profile with linear 10 fs intensity growth, 40 fs constant intensity and 10 fs linear intensity decrease. The spatial profile was Gaussian, laser beam diameter (FWHM) was set to 4 $\rm{\mu m}$ at focus. Laser was focused into plane coordinates (0,0). The simulation box size was set to $31\times31\lambda^2$ and to $34\times31\lambda^2$ for targets without/with preplasma, which means that it contains $1488\times1488$ and $1632\times1488$ cells, $4.4\times10^{7}$ and $7.8\times10^{7}$ macroparticles, respectively. The initial temperature of macroparticles was 1~keV, the macroparticles were frozen before (pre-)plasma interaction with the laser in the simulations. The cell size was set to be about 2/3 of the skin depth in order to resolve relevant physics. Units in the figures are either specified explicitly, or following reference values are used for the normalization: electric field $E_r=m_ec\omega/e$, magnetic field $B_r=m_e\omega/e$, momentum $p_r=m_ec$, force $F_r=m_ec\omega$, where $m_e$ is mass of an electron, $c$ is speed of light, $\omega$ is laser frequency, $e$ is elementary charge.  
	
	\begin{figure}
		\includegraphics[width=.45\textwidth]{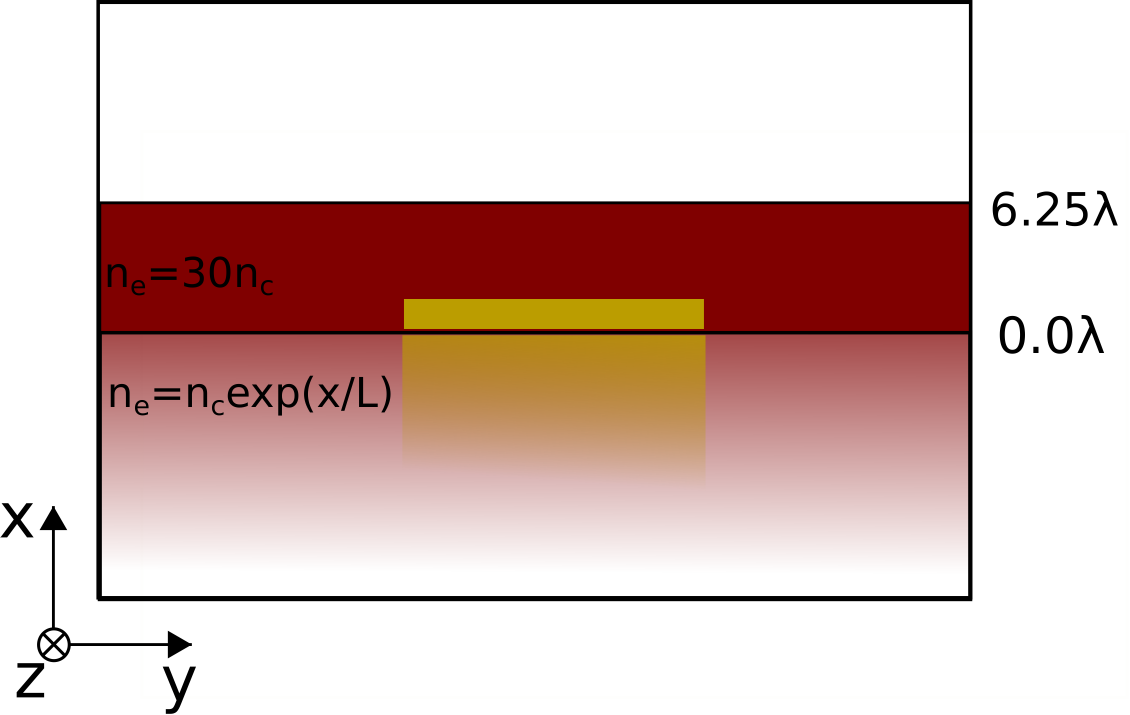}
		\caption{\label{fig:sim_oblast} The position of overdense target with preplasma in front of it. Yellow color shows the position of test particles.}
	\end{figure}
	
	This work vastly uses particle tracking diagnostics implemented in Smilei together with test particles. Test particles do not contribute to charge and current densities, but the code calculates their trajectories from EM fields at their location. Particle tracking diagnostics saved position, momenta, and fields at the position of particle. At simulations with steep density profile, test electrons were put at the vacuum-plasma interface into the strip of width $3\lambda$ ($x=\left\langle 0, 3\lambda \right\rangle$) irradiated by the laser beam where hot electrons are expected to be generated. For simulations with preplasma, test electrons are present along the entire length of preplasma in $x$ direction. In $y$ direction, test electrons were placed inside the region $y=\left\langle -6\lambda,6\lambda \right\rangle$ both in simulation with and without preplasma. Their location is displayed by yellow color in Fig. \ref{fig:sim_oblast}. All particles were tracked from the beginning of the simulation. 
	
	Hot electron population can be easily distinguished from the target bulk electron population in energy distribution function. Hot and cold electron temperatures can be assigned to both populations with Boltzmann energy distribution \cite{liseykina_dist,tikhonchuk_ionacc}. Therefore, the minimal energy $E_{min}$ of hot electron population can be defined at the place where the slope of high energy tail changes in energy distribution function. During the post-processing, only those particles that were inside the overdense part of the target while satisfying the condition $E>E_{min}$ were analyzed.  

    \section{Steep density gradient}
    When laser pulse interacted with target with steep density gradient, the interaction started at time 17.5$T$ and finished at time 40$T$, where $T$ is laser period. 

	\subsection{Intensity $10^{18}~\rm{W/cm^2}$}
\begin{figure}
		\includegraphics[width=1\textwidth]{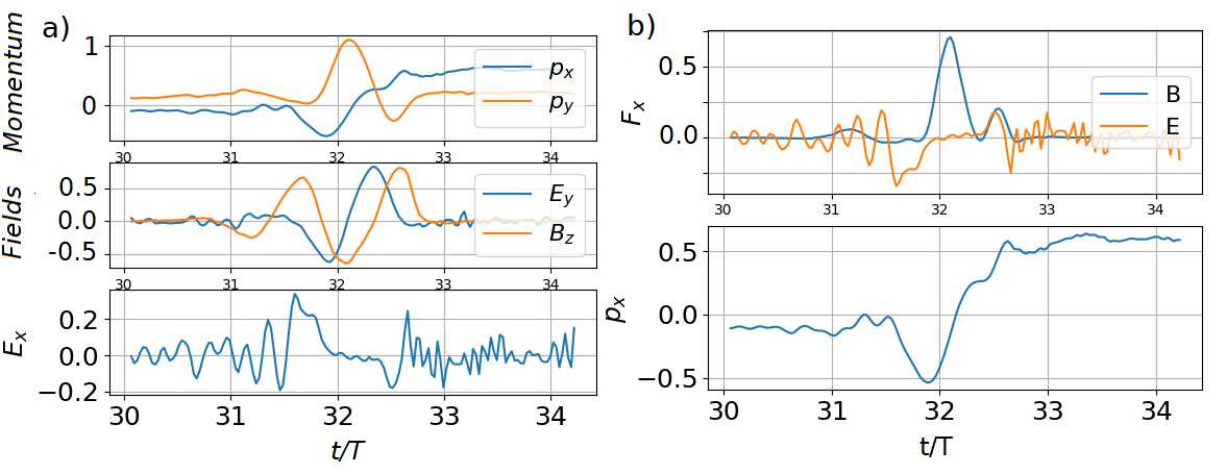}	
		\caption{(a) The evolution of momentum and fields acting on an electron during the acceleration process by normally incident laser pulse with intensity $10^{18}~\rm{W/cm^2}$. (b) The time evolution of forces acting on the electron. At the beginning, the electrostatic force (E) pushes the electron to the area in front of the target and consequently the magnetic part of Lorentz force (B) causes the acceleration inside the target.}
		\label{fig:18-0-fields}
		\label{fig:18-0-forces}
	\end{figure}

    Hot electron acceleration mechanism observed in our simulation results can be described as follows. In Fig. \ref{fig:18-0-fields}, the evolution of tracked electron quantities during the acceleration is depicted. This evolution of momentum and fields acting on particle represents qualitative behaviour common for all hot electrons. At the beginning of each acceleration process, the growth of momentum in a direction out of the target ($p_x<0$) occurs. This happens due to the effect of electrostatic field pulling the electron out to the vacuum as can be seen in Fig. \ref{fig:18-0-fields} a) at time around 31.5$T$. $p_x$ is negative at the same time when $E_x$ component of electric field is positive at the position of particle, which results in electric component of Lorentz force acting on electron in direction out of the target, see Fig. \ref{fig:18-0-forces} b). Magnetic component of Lorentz force (B) has no impact on electron being pulled out to the vacuum. The electrostatic field on the boundary of plasma and vacuum is created because of oscillation of electrons around immobile ions (on the time scale of a laser period as they have much larger mass) with frequency of 2$\omega$, where $\omega$ is the laser frequency. This ejection of electrons into the vacuum by electrostatic field was previously described in Refs.~\onlinecite{debayle,cai_jxb}.

\begin{figure}
		\includegraphics[width = .455\textwidth]{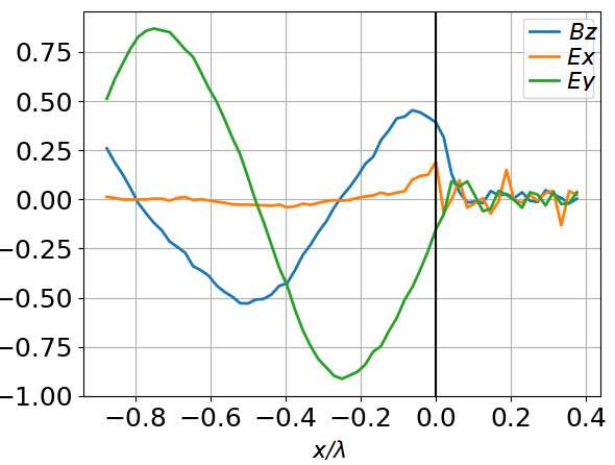}
		\caption{Standing wave which was created after the reflection of laser pulse with intensity $10^{18}~\rm{W/cm^2}$ from the target without preplasma at time 33.4$T$. Fields are normalized by reference fields: electric field $E_r=m_ec\omega/e$, magnetic field $B_r=m_e\omega/e$. It is visible that the standing wave is shifted because of finite conductivity of the target and consequent skin effect. } 
		\label{fig:18-0-sw}
	\end{figure}

\begin{figure}
	\includegraphics[width=.95\textwidth]{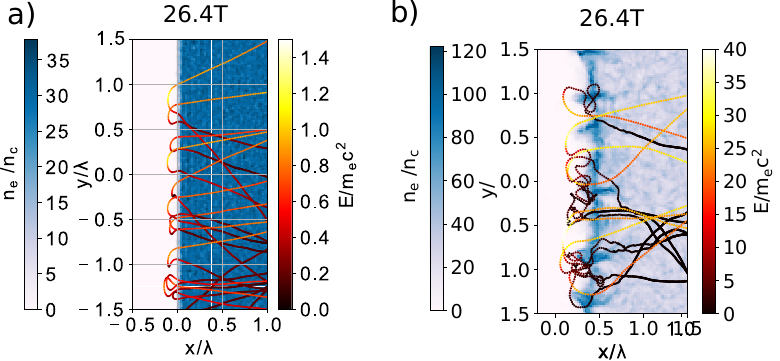}
	\caption{Density of the target during the interaction with laser pulse with intensity $10^{18}~\rm{W/cm^2}$ (a) and $10^{21}~\rm{W/cm^2}$ (b) with typical electron trajectories. It can be seen that EPB is completely unperturbed for lower intensity and electrons are accelerated immediately after being pulled out of the target. Higher intensity pushes the EPB inside the target and electrons oscillate chaotically before being accelerated.  }
	\label{fig:traj_dens_sharp}
    \end{figure}

	When electron is already in the vacuum, it interacts with the standing EM wave created in front of the target and depicted in Fig. \ref{fig:18-0-sw}. Electric component of the standing wave $E_y$ causes that electron gains momentum in the direction along the target surface. Acceleration in the $x-$direction starts at the moment when the magnetic field perpendicular to the plane of incidence $B_z$ changes its sign as shown in Fig. \ref{fig:18-0-fields} a) in time shortly before 32$T$. At that time, the product of $v_yB_z$ is negative, so the Lorentz force acts on a negatively charged electron in a positive direction towards the target. Relativistic values of velocity $v_y$ lead to a relatively high strength of the magnetic part of Lorentz force in perpendicular direction to the standing wave $B_z$ and $E_y$ fields orientation, i.e., perpendicularly to the target surface.    
The process of acceleration ends at the moment when particle crosses the plasma boundary and doesn’t feel the standing wave field. The typical trajectories of accelerated electrons can be seen in Fig. \ref{fig:traj_dens_sharp} a).
	
	According to how the acceleration process was described, it is possible to assume that the standing wave structure has the most significant impact on the acceleration process. Electrons immediately react to changes of the perpendicular electric field $E_y$. It is also possible to see that $B_z$ and $p_y$ values are shifted by $\pi$ in phase and they have the same period. Thus, the force proportional to $v_yB_z$ is always positive. Since the electric field is harmonic in time, the phase shift between $p_y$ and $E_y$ is $\pi/2$. The phase shift between $E_y$ and $B_z$, which is also $\pi/2$ (see Fig. \ref{fig:18-0-sw}), causes already mentioned $\pi$ phase shift of $p_y$ and $B_z$. The action of the magnetic part of Lorentz force thus twice a period accelerates electrons into the target. 

\begin{figure}
		\includegraphics[width=1\textwidth]{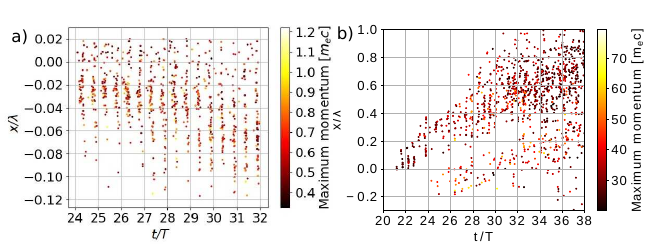}
		\caption{Times and positions when electron acceleration towards the target started during the interaction with normally incident pulse of intensity (a) $10^{18}~\rm{W/cm^2}$ and (b) $10^{21}~\rm{W/cm^2}$ on target with steep density gradient.}
		\label{fig:18-0-min}		
		\label{fig:21-0-min}
	\end{figure}

All hot electrons were accelerated in the region between the first node of the magnetic field at distance $\approx 1/4\lambda$ in front of the target and EPB, see Fig. \ref{fig:18-0-sw}. 
In Fig. \ref{fig:18-0-min}, points on the graph show the time and the position of electrons at the beginning of acceleration. 
The color gives information about the maximal longitudinal momentum of electrons after the acceleration. 
The moment of the beginning of the acceleration is here defined as the moment when $p_x$ reaches its minimal value. 
The acceleration twice a period typical for $\vec{j}\times\vec{B}$ heating can be observed from vertical bunches repeating every half-period. 
As mentioned before, acceleration starts at the moment when the magnetic field changes its sign. 
This happens at the same time at positions of all electrons in front of the target twice a period. 
That is the reason why all electrons within one bunch start to be accelerated at the same time independently from its position.

Note that minimal $x$-position for each accelerated electron is below its position depicted in Fig. \ref{fig:18-0-min} a) as the longitudinal momentum $p_x$ is still negative in the beginning of electron longitudinal acceleration towards the target. 
The position of electrons is about $x=-0.1\lambda$, whereas the minimum position is at around $x=-0.15\lambda$ and the electrons still does not move behind magnetic field node at $x \approx -1/4\lambda$. 
One can expect that the position of an electron during the acceleration determines its longitudinal momentum increase after the acceleration which depends on transverse electric field $E_y$, transverse magnetic field $B_z$, and electron acceleration time.
The thickness of the acceleration region in front of the target should be proportional only to the strength of the longitudinal field $E_x$ pulling electrons out into the vacuum. 
In semi-analytical model \cite{debayle} assuming infinite ion density gradient with immobile ions, the strength of such $E_x$ field is theoretically fully determined by laser pulse amplitude and plasma density. 
However, the target expands during the interaction with main laser pulse and due to non-zero temperature of plasma before the main pulse interaction (electron heating by laser pedestal or by picosecond slope of the pulse rising slowly compared to an ideal Gaussian pulse \cite{Wagner2016}). 
Therefore, the width of the acceleration region strongly depends on the real value of steep plasma density gradient and evolves during the interaction (increases with decreasing ion density gradient in time). That is also apparent in Fig. \ref{fig:18-0-min} a) where more than 90 percent of the electrons start to be accelerated in the $x$-direction at position $0> x > -0.05\lambda$ at time around 24$T$  but at wider region $0> x > -0.1\lambda$ at later time around 32$T$.
 
\subsection{Intensity $10^{21}~\rm{W/cm^2}$}
	
When laser pulse with intensity $10^{21}~\rm{W/cm^2}$ was incident on target with steep gradient, interaction regime significantly changed from idealized scenario. The strong radiation pressure deformed the target and bored a hole in it. Consequently, initially step-like interface was strongly disrupted and interaction region consisted of local density, electric and magnetic field disturbances. Curved shape of overdense plasma boundary caused that reflected wave was strongly non-uniform. Therefore, the standing wave in front of the target was far from ideal. Due to many violations in the interaction geometry, hot electrons were accelerated and decelerated in front of the target before they crossed the boundary of overdense foil, as can be seen in Fig. \ref{fig:traj_dens_sharp} b). The impact of electric field is negligible, longitudinal motion of electron is determined by the magnetic component of Lorentz force. 

In Fig. \ref{fig:21-0-min} b), one can see the starting points of acceleration defined in the same manner as in the previous case. 
The EPB being pushed can be observed in the left top corner. 
It can be estimated from the figure that the plasma layer was pushed with velocity $\sim 0.08c$. 
The value of EPB velocity agrees with the value obtained from densities. 
Since electrons are mostly accelerated close to the boundary, the starting point of acceleration moves with it. 
Firstly, it moves quickly up to the moment when an equilibrium is established between the electrostatic pressure generated by compressed electron cloud and the radiation pressure of laser beam on the time scale of laser period \cite{klimo_hb}.
Then, the hole boring process continues both in electron and ion densities during the whole interaction \cite{Robinson2009}. 
This can be seen in the figure up to the time around 31$T$ when the points are more scattered along $x$-position due to the increasing difference in EPB deformation along $y$-axis. 

The estimated EPB deformation at the first moment of the pressures equilibrium \cite{klimo_hb} $\Delta_{init} = \sqrt{4 \epsilon_0 I / c}/(e n_e) \approx 130~\rm{nm}$ and the estimated hole boring velocity \cite{Robinson2009} $u_{hb}/c = \sqrt{\Xi}/(1+\sqrt{\Xi}) \approx 0.06$, where $\Xi = I / (m_i n_i c^3)$, agrees well with our observations in the simulation. According to these analytical estimates, one can expect less pronounced deformation of the target by strong radiation pressure with increasing target density (e.g., using metal foils). Note that EPB deformation can be seen in Fig. \ref{fig:traj_dens_sharp} b) when the deformation exceeds the initial value $\Delta_{init}$ due to hole boring process after the initial EPB deformation. 

The width of the acceleration region is substantially larger along $x$-direction compared with $10^{18}~\rm{W/cm^2}$ case. 
Although the standing wave structure is disrupted, it is apparent that some electrons are accelerated beyond the first node of the magnetic field from EPB in this non-ideal standing wave. 
Two strips of points where electrons are started to be accelerated are perceptible in Fig. \ref{fig:21-0-min} b). 
The strip located closer to EPB is between EPB and the first node, the other strip is at some distance from the first magnetic field node where the combination of instantaneous transverse electric field $E_y$ and magnetic field $B_z$ is favourable for electron acceleration towards the overdense plasma. 
	
The initial density of overdense target $n_e \approx 30 n_c$ is relatively low with regard to the commonly presented threshold for relativistic transparency $a_{th} = n_e/n_c$ which is close to the pulse amplitude $a_0 \approx 22$ for a given laser intensity. 
However, this threshold is essentially increased to $a_{th} \approx 0.65(n_e/n_c)^2$ for $(n_e/n_c) >> 1$ in the case of non-uniform plasmas due to the action of the ponderomotive force which pushes electrons into the plasma creating a strong peaking of the plasma electron density \cite{Cattani2000}. 
Moreover, the total pulse length is 60~fs in our case, which is not enough for substantial plasma expansion and lowering the plasma density. 
Thus, we are still far from relativistically induced transparency regime during the whole interaction even with the lowest solid density foil made from hydrogen.

	\section{Effects of preplasma}
	
	In the study of the interaction of laser pulse with overdense target slightly evaporated by the pulse pedestal, preplasma with exponentially increasing density $n_e = n_c\times\exp(x/L)$ for $x\leq0$ and scale length $L=3 \rm{\mu}$m was assumed in front of the overdense ionised foil located at $0<x<5~\mu$m , see Fig. \ref{fig:sim_oblast}. Therefore, the distance between the simulation box boundary where laser pulse enters and the overdense plasma had to be increased and laser pulse started to reflect from overdense thin foil at time 20$T$ and stopped the interaction at time 42.5$T$. 
	\subsection{Intensity $10^{18}~\rm{W/cm^2}$}
	The presence of preplasma in front of the target significantly changed the interaction regime. The most of electrons were accelerated in underdense preplasma, not in the proximity of overdense plasma boundary like in the previous case with steep gradient. 
	
	\begin{figure}
		\centering
		\includegraphics[width=.85\textwidth]{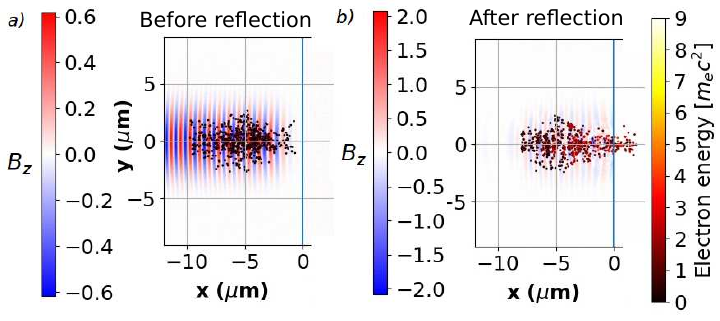}
		\caption{(a) Electrons oscillating perpendicularly in the preplasma at the position of laser pulse with intensity $10^{18}~\rm{W/cm^2}$ propagating towards the overdense part of the target. (b) Electrons on the right from the reflected wave's wavefront at $x=-6 ~\rm{\mu}$m are red, which means that they were already accelerated towards	the overdense part of the target while those on the left are still oscillating	in the field of incident wave. }
		\label{fig:pre-18-0-sw}
	\end{figure}
	
\begin{figure}
		\includegraphics[width=.45\textwidth]{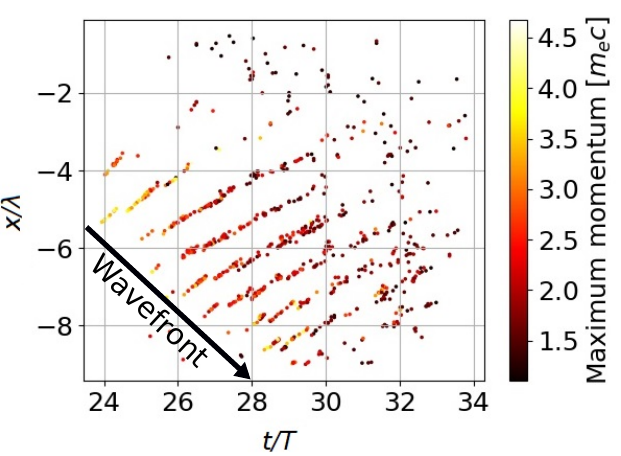}
		\caption{Times and positions when electron acceleration towards the target started during the interaction with the pulse with intensity $10^{18}~\rm{W/cm^2}$ on target with preplasma. Black line shows the position of the wavefront of the reflected wave from overdense plasma layer.}
		\label{fig:pre-18-0-min}
	\end{figure}

	Acceleration process can be described as follows. While laser pulse propagates through the preplasma, electrons oscillate perpendicularly to the pulse propagation direction due to the effect of oscillating electric field. When laser wave is reflected, it starts to propagate in the direction out of the target and the superposition of reflected and incident wave creates a non-ideal standing wave. At the moment when reflected wave reaches the position of oscillating electron, it is accelerated inside the target as can be seen in Fig. \ref{fig:pre-18-0-sw}. In a), electrons are not accelerated yet (black color of dots) because laser pulse still propagates towards the overcritical target surface. In b), the wavefront of reflected wave is located at around  $x=-6~\rm{\mu m}$ and the standing wave is already created at positions $x>-6~\rm{\mu m}$. The color of electrons indicates that they are already accelerated into the target. It can be seen that electrons on the left from the wavefront were not accelerated yet because standing wave is just about to be formed in that region. This clearly shows the impact of standing wave on electron acceleration. The fact that electrons are not accelerated before the reflected wave reaches their position is demonstrated by the maximum distance of acceleration from the target in Fig. \ref{fig:pre-18-0-min}. The maximum distance of acceleration increases with time and corresponds to the position of reflected wave's wavefront. The position of reflected wavefront is approximately shown by the black line.
	
	\begin{figure*}
		\centering
		\includegraphics[width=1\textwidth]{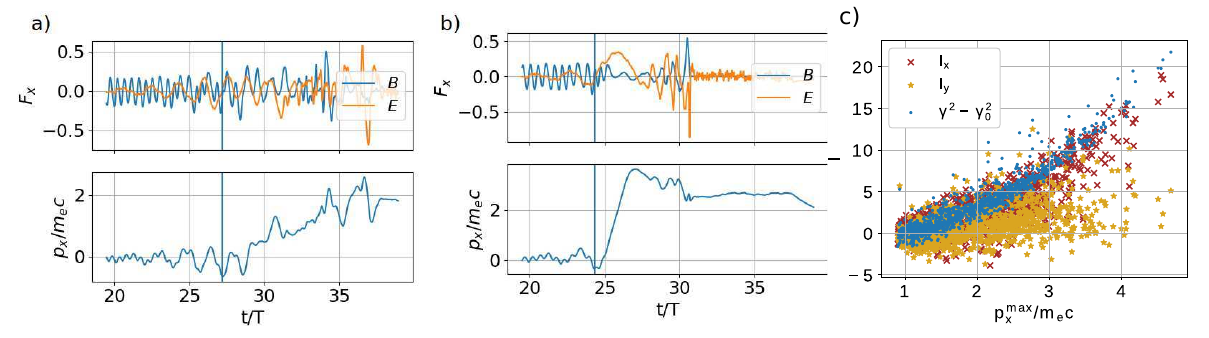}
		\caption{Evolution of forces acting on electron in during the acceleration and $p_x$ of the electron when laser with intensity $10^{18}~\rm{W/cm^2}$ is incident on the target with preplasma. Vertical line indicates the moment of acceleration due to our definition. In (a), electron was accelerated by the stochastic effect of electric and magnetic components of Lorentz force. In (b) electron was accelerated by the longitudinal electric field. (c) The contribution of electric field components to the acceleration at the moment of maximum $p_x$. }
		\label{fig:pre-18-0-fx}
	\end{figure*}

	The evolution of electron momentum and Lorentz force components acting on it are depicted in Fig. \ref{fig:pre-18-0-fx}. The blue vertical line defines the starting point of acceleration used in Fig. \ref{fig:pre-18-0-min}. It can be seen in Fig. \ref{fig:pre-18-0-fx} a) that the starting point of acceleration is only a rough estimate since more $p_x$ minima are present close to the vertical line, however it is sufficient for our analysis. Fig. \ref{fig:pre-18-0-fx} illustrates two boundary cases of acceleration scenarios  observed for all tracked electrons. 

\begin{figure*}
		\includegraphics[width = .55\textwidth]{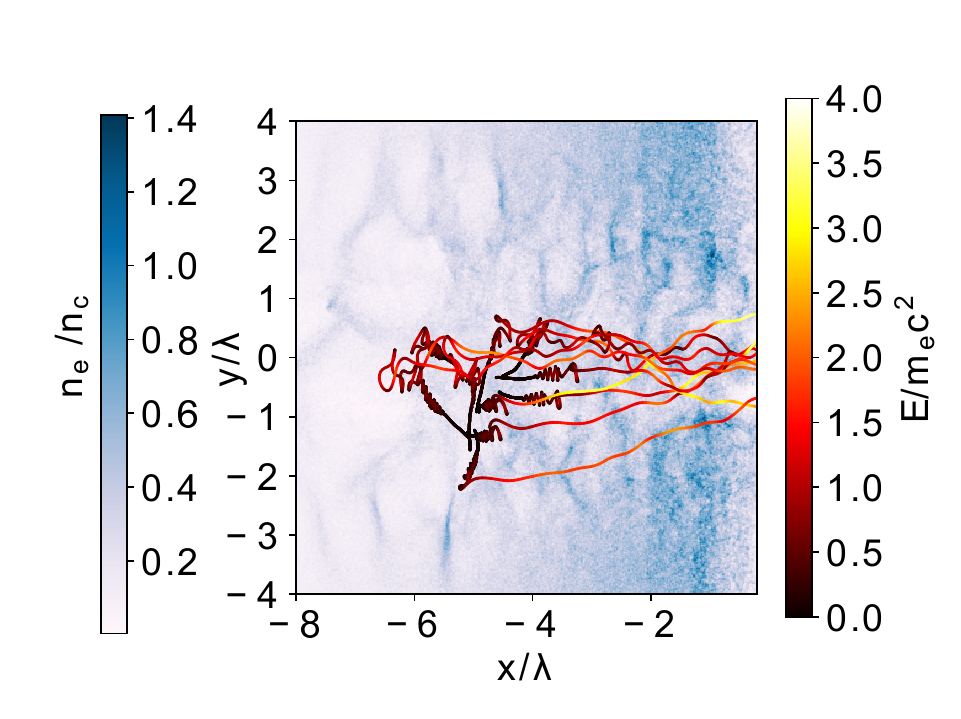}
		\caption{Typical electron trajectories during the acceleration after the incidence of pulse with intensity $10^{18}~\rm{W/cm^2}$ on target with preplasma. }
		\label{fig:traj_dens18}	
	\end{figure*}

	Many electrons were accelerated in a way similar to Fig. \ref{fig:pre-18-0-fx}a). The negative momentum gain is again present followed by the acceleration towards the target, but while electron propagated through preplasma it was periodically accelerated until it reached the overdense plasma at time approximately 37$T$. The final momentum gain was the result of several pushes by the standing wave and longitudinal $E_x$ while electron propagated towards the target. On the other hand, some electrons had very similar $p_x$ evolution during the acceleration process as electrons accelerated in the target with steep gradient, see Fig. \ref{fig:pre-18-0-fx}b). At time around 25$T$, electron gains momentum in the direction away from the target and right after it is accelerated towards the target. Here, the initial negative momentum gain occurs in preplasma and does not mean pulling the electron out of the overdense plasma layer. The second difference is that mostly longitudinal electric field component was responsible for the acceleration, not the standing wave field. However, the most of electrons were accelerated by the combination of several strong electric field pushes similar to Fig. \ref{fig:pre-18-0-fx}b) and by the chaotic accelerations and decelerations similar to Fig. \ref{fig:pre-18-0-fx}a). Typical electron trajectories are depicted in Fig. \ref{fig:traj_dens18}
	
To quantify the effect of mechanisms, the contribution of electric field components was expressed using formula $\gamma^2(t)-\gamma_0^2 = -2\int_{t_0}^t E_xp_xd\tau-2\int_{t_0}^t E_yp_yd\tau-2\int_{t_0}^t E_zp_zd\tau = I_x+I_y+I_z$, where $\gamma_0$ is the initial gamma factor of accelerated electron. The value of integral $I_x$ represents the contribution longitudinal electric field which can be present because of charge separation and laser self-focusing. The integral $I_y$ represents the contribution of background transverse field and the oscillating field of laser pulse which was consequently bend into $x$-direction by the magnetic field. In chosen geometry the value of $I_z$ can be neglected.The value of integrals was evaluated at the moment when electron reached its maximum in $p_x$, see Fig. \ref{fig:pre-18-0-fx}c). 
	
The effect of electric and magnetic field on electron acceleration is comparable for electrons with lower maximal momentum, namely $p_x^{max}<2.5m_ec$. This is in agreement with acceleration process described in Fig. \ref{fig:pre-18-0-fx}a), where the combination electric and magnetic Lorentz force contributes to the final momentum. Electrons with the highest energies were accelerated mostly by the longitudinal electric field, see Fig. \ref{fig:pre-18-0-fx}b). 

The mechanism responsible for acceleration by the transverse oscillating field is stochastic heating. The acceleration happens in the field of standing wave and the Lyapunov exponent\cite{lieberman_stoch} for accelerated electrons is positive. The value of Lyapunov exponent was calculated in a following way. Pairs of randomly selected hot electrons were created and the Lypunov exponent was calculated for all pairs according to Ref. \onlinecite{lieberman_stoch}. The initial time $t_0$ was selected as a time when reflected laser wave reached the position of the electron. Afterwards, the sequence of Lyapunov exponents was created as a function of initial electron pair distance in phase space $d(\vec{x_0},t_0)$. The final value of Lyapunov exponent was obtained from the sequence as a limit for $d(\vec{x_0},t_0)\to 0$. 

The positive value of Lyapunov coefficient demonstrates the stochasticity of electron motion. To confirm this we performed the simulation with counter-propagating pulses with intensity $10^{18}~\rm{W/cm^2}$, where only test particles were present. The effect of background field does not affect the electron motion in such simulation. The Lyapunov exponent of electron motion in the standing wave created by the counter-propagating pulses was positive and the acceleration was present with maximum $ p_x$ exceeding $m_ec$. Only $I_y$ was non-zero during this acceleration which means that electron's oscillations in transverse direction were turned into positive x-direction. Based on just mentioned observation we conclude that mechanism responsible for electron acceleration in PIC simulations is the combination of stochastic heating and longitudinal electric field. 
	
	\subsection{Intensity $10^{21}~\rm{W/cm^2}$}\label{ch:e21pre}
	
	\begin{figure}
	\includegraphics[width=.45\textwidth]{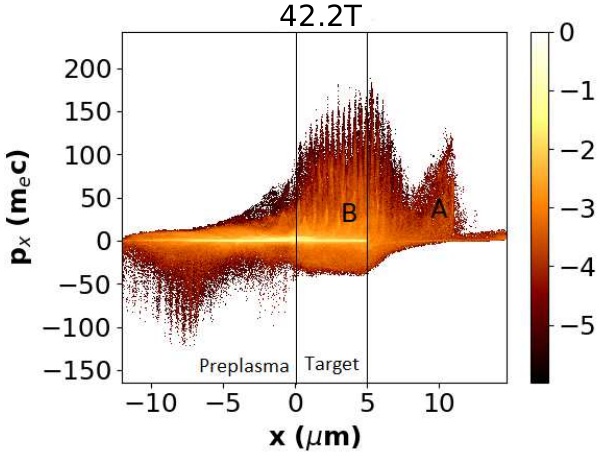}
	\caption{Phase space of electrons after	the incidence of laser pulse with intensity	$10^{21}~\rm{W/cm^2}$ on the target with preplasma. Color	shows the relative amount of electrons where numbers on colorbar corresponds to the exponent on the logarithmic scale. }
	\label{fig:pre-21-0-xpx}
	\end{figure}

	\begin{figure*}
	\centering
	\includegraphics[width=1\textwidth]{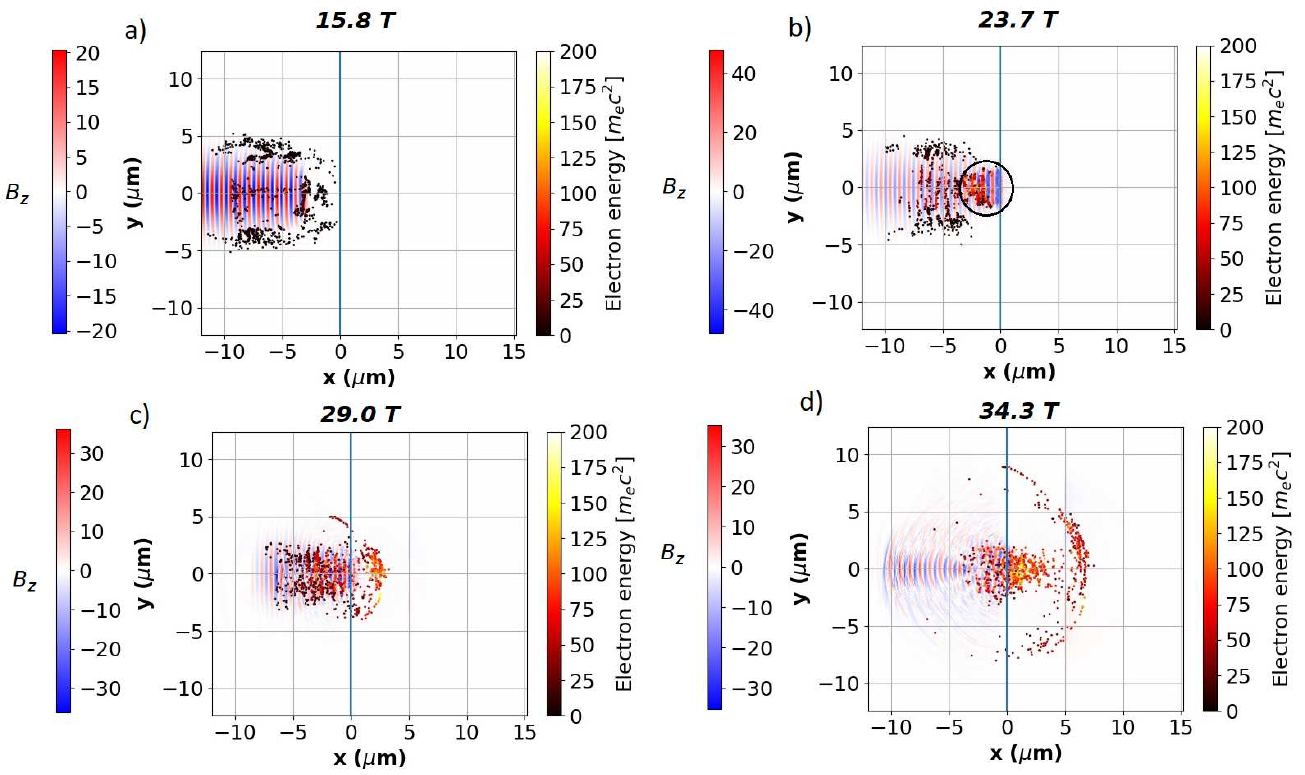}
	\caption{Positions and energies of hot electrons during the acceleration after the incidence of laser pulse with intensity $10^{21}~\rm{W/cm^2}$ on the target with preplasma. (a) Travelling wave repels electrons; (b) Electrons are injected into the pulse where standing wave is created. Electrons from population A can be marked by the black circle. Electrons outside the circle belong to the population B. (c)-(d) Electrons are accelerated in the field of standing wave. }
	\label{fig:pre-21-0-scatter}
    \end{figure*}

	Higher intensity of laser pulse brought several new phenomena into the interaction compared to the case with intensity $10^{18}~\rm{W/cm^2}$ while the dominant impact of standing EM wave was still present. Two hot electron populations can be distinguished in Fig. \ref{fig:pre-21-0-xpx}. The first one (population A) is located at $x$-axis position around 10 $\rm{\mu m}$ and another one is spread between $x=0$ and $7$ $\rm{\mu m}$ (population B). The rough estimate is that 45$\%$ of hot electrons belong to the population A and 55$\%$ of hot electrons belong to the population B. The population A doesn't seem to be formed of separate electron bunches as expected for electrons accelerated by normally incident pulse. The distances between bunches were smoothened when electrons crossed the rear side of the target at $x=5\rm{\mu m}$ and were clearly visible at earlier times when they were still located in front of or inside the target. The distance between electrons in the population B is $\lambda/2$ since they still didn't cross the rear side. Distances between the bunches are identical as in the case of $\vec{j}\times\vec{B}$ heating even though electrons were not accelerated at overdense plasma boundary but rather in preplasma.

	The acceleration process can be described by Fig. \ref{fig:pre-21-0-scatter}. Four figures describe the propagation of laser pulse in preplasma towards the overdense plasma layer with hot electrons positions. 

	In a) laser pulse is propagating through preplasma. Electrons are either pushed forward in front of the pulse or to the sides of the pulse by the ponderomotive force. The color of electrons shows that they were not accelerated yet because the laser pulse was still propagating towards the target and standing wave was not created yet.

	In b) the splitting of electrons into two populations can be already observed. The population A is located closest to the overdense layer marked by the blue vertical line. It is marked inside the black circle. This shows that the population A is created mostly by the electrons that are pushed in front of the pulse. All electrons that are outside the circle belong to the population B. They are plotted either as a black unaccelerated dots on the sides of the pulse or they are already starting to form lines perpendicular to the laser propagation direction by being attracted to the center of the beam with high intensity laser field. 

	In c) the population B is already accelerated and is located inside the target. The blue line showing the critical density layer roughly separates population A from population B. All electrons that were previously on the sides of laser pulse were attracted to the region of laser pulse and they are accelerated towards the overdense part of the target. 

	In d) the most of electrons from the first population already crossed the rear side of the foil and majority of electrons from population B is propagating inside the target while the small part still interacts with the electromagnetic field in front of the target.

\begin{figure}
		\includegraphics[width = .45\textwidth]{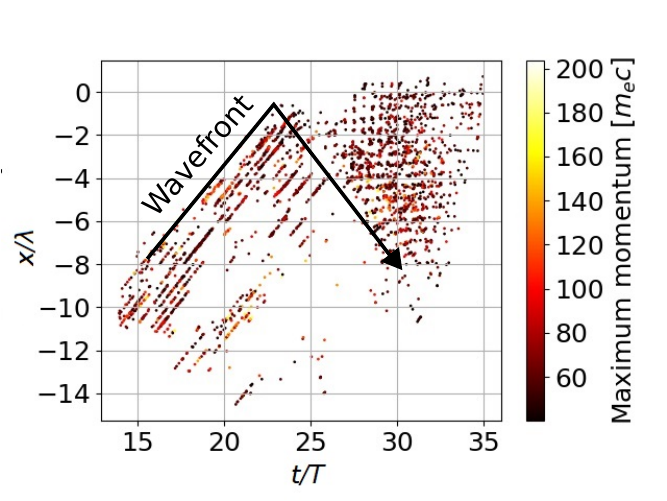}
		\caption{Times and positions when electron acceleration towards the target started during the interaction with normally incident pulse of intensity $10^{21}~\rm{W/cm^2}$ on target with preplasma. Black line shows the position of the wavefront of the laser wave before and after the reflection from overdense plasma layer.}
		\label{fig:pre-21-0-min}	
	\end{figure}

	Two populations (A and B) are also easily observed in Fig. \ref{fig:pre-21-0-min} where initial positions and times of the start of acceleration are shown. The initial time of acceleration is again defined as a time when $p_x$ has a minimum value. The black arrow represents the position of laser pulse wavefront. The point (23.5$T$,0$\rm{\lambda}$) represents the time when constant part of the laser pulse is reflected from the overdense region. The arrow coming from this point separates two hot electron populations from each other. On the left, straight lines parallel to incident wave's wavefront correspond to times when electrons were pushed by the propagating laser pulse towards the target. Electrons depicted on the right were not pushed by the propagating beam. Instead, they were injected to the high-intensity part of the laser pulse from the sides and afterwards accelerated similarly to the case describing pulse with intensity $10^{18}~\rm{W/cm^2}$. Again, the maximum distance of acceleration at certain time is constrained by the position of wavefront of reflected wave. For those electrons the standing wave seems to be inevitable condition for acceleration to occur.

	\begin{figure*}
		\includegraphics[width = 1\textwidth]{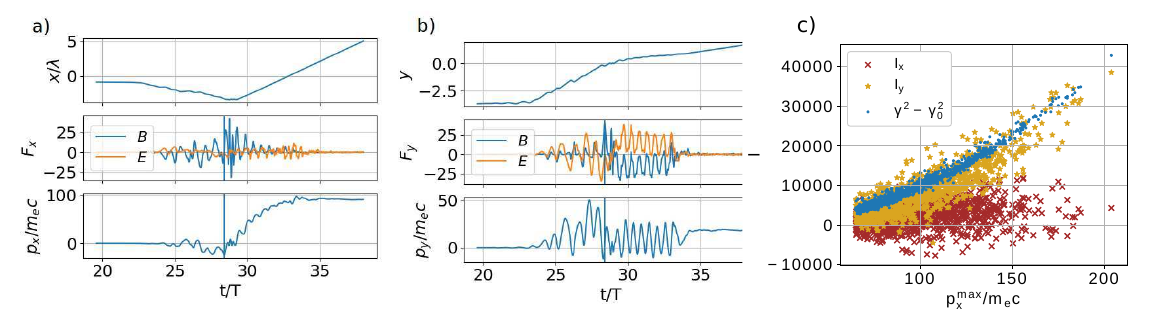}
		\caption{Evolution of forces acting on an electron	during its acceleration, position and momentum of the electron when laser with intensity $10^{21}~\rm{W/cm^2}$ is incident on the target with preplasma. Vertical line indicates the moment of acceleration according to our definition. (a) $x$ components; (b) $y$ components; (c) The contribution of the electric field components at the time of $p_x$ maximum and the increase of gamma factor. }
		\label{fig:pre-21-0-f}	
	\end{figure*}

		\begin{figure*}
		\includegraphics[width = .55\textwidth]{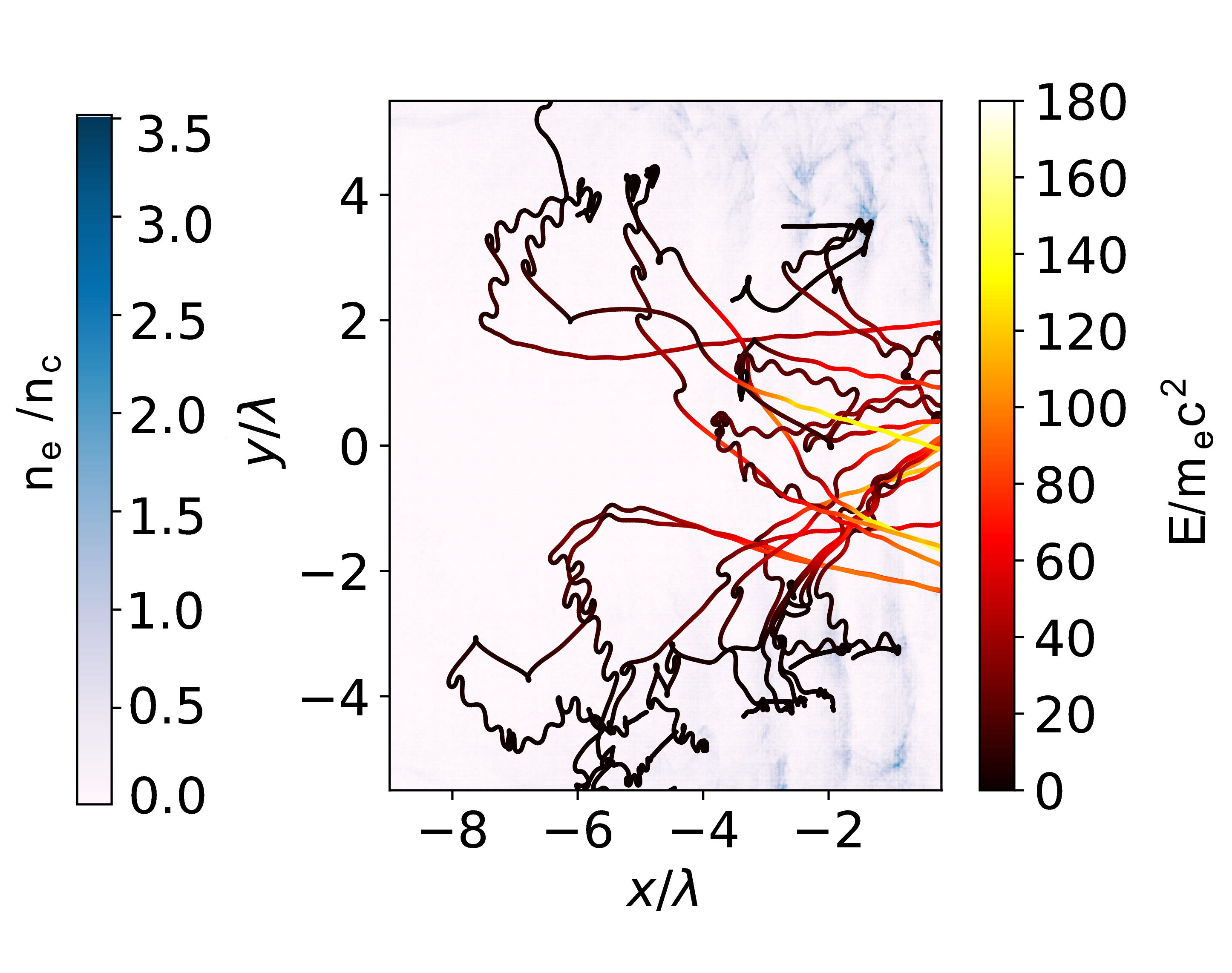}
		\caption{Typical electron trajectories during their acceleration after the incidence of pulse with intensity $10^{21}~\rm{W/cm^2}$ on the target with preplasma. The repulsion of electrons on the sides of laser pulse can be observed from electron density. Electrons oscillate on the sides of laser pulse (black part of trajectory) and they are injected into the beam axis region where the acceleration happens afterwards (red/yellow part of trajectory).}
		\label{fig:traj_dens21}	
	\end{figure*}

	After a look on $y$ position of typically accelerated electron from the population B in Fig. \ref{fig:pre-21-0-f} b), it can be seen that electron was moving towards the centre of laser pulse before it was accelerated. This electron movement along the target surface to the centre of laser pulse is caused by the $p_y$ momentum component oscillating in positive values before the blue vertical line showing the starting time of acceleration used in Fig. \ref{fig:pre-21-0-min}. Such injection into the high-field regions occurs because of the presence of standing wave and plasma channel and is further analyzed in section \ref{ch:sw}. After electron injection into the high field region, it is accelerated towards the target, see Fig. \ref{fig:pre-21-0-f} a) where the electron was consecutively accelerated during 5 laser periods while it propagated towards the target in the standing wave. Additive impact of several pushes from the standing wave while it propagated in preplasma allowed electron population to be heated to temperature close to ponderomotive scaling \cite{wilks_abs} $T_h=((1+a_0^2)^{1/2}-1)m_ec^2$, where $T_h$ is hot electron temperature in eV. Namely, the temperature of hot electrons observed in the simulation was $8.3$ MeV, using Boltzman distribution. Temperatures of hot electrons were higher than those predicted by the Beg's scaling \cite{beg_scaling}. For the case of lower intensity, hot electrons even exceed the ponderomotive scaling. The temperature of hot electrons accelerated in the target with preplasma exceeds the temperature of electrons accelerated by the pulse with same intensity in the target with steep density gradient. This was observed for both laser intensities, see Table \ref{tab:temps}.

	\begin{table}
		\caption{\label{tab:temps}Comparison of hot electron temperatures for different scenarios fitted using Boltzman distribution compared with ponderomotive scaling $T_{pond}=((1+a_0^2)^{1/2}-1)m_ec^2$  }
		\begin{ruledtabular}
			\begin{tabular}{ccccc}
			&T[keV] &T[keV]  & T[keV] & \\
				I[$\rm{Wcm^{-2}}$]	& steep density& preplasma &preplasma& $\rm{T_{pond}}$[keV]\\
				&  p-polarization& p-polarization &s-polarization& \\
				\hline
			$10^{18}$ & 24 & 250 & 150 &96\\
				$10^{21}$ & 4800 & 8300 & 7200 &10538\\
			\end{tabular}
		\end{ruledtabular}
	\end{table}

To determine the acceleration mechanism, we have calculated the contribution of electric field components and compared the values at the moment of $p_x$ maximum, see Fig. \ref{fig:pre-21-0-f} c). The integral $I_x$ corresponds to the contribution of longitudinal electric field. This is usually not significantly present during the normal incidence. However, the radiation pressure deformed the overdense target surface into parabolic-like shape which focused the laser pulse and turned the $E_y$ component of laser field into $x$-direction. The relativistic self-focusing and charge separation are also present but it is not possible to distinguish between the possible $E_x$ field sources because of simulation complexity. The integral $I_y$ corresponds to the contribution of standing wave and background transverse field. Since the difference between electron temperatures for different polarization were previously observed \cite{Stark2017}, we performed simulations of target with preplasma also with s-polarized pulse. They showed that the interaction with p-polarized pulse leads to higher electron temperature. We propose the following explanation. When electrons are accelerated by the standing wave, they oscillate in either positive or negative values of $p_y$ while magnetic field of the standing wave transfers the momentum into $x$-direction. When the background field of plasma channel is present, the channel electric field can increase the mean value of $p_y$ oscillations. Therefore, the higher transverse momentum can be transferred into $x-$direction by magnetic field which results in higher electron momentum and temperature. When oscillating electric field is oriented out of simulation plane in s-polarized pulse, the effect of background electric field on electron energy is negligible and results in lower electron temperatures as expected. The positive value of Lyapunov exponent for electron trajectories indicates that the mechanism of electron acceleration corresponding to $I_y$ is stochastic heating. 

The effect of longitudinal electric field and stochastic heating on electron acceleration is comparable for electrons with lower maximum momenta ($p_x^{max}<120~m_ec$). On the other side, stochastic heating is the dominant acceleration mechanism for the most energetic electrons. Typical electron trajectories are depicted in Fig. \ref{fig:traj_dens21}. At the beginning, electrons are located on the sides of the laser pulse	because they were repelled by the ponderomotive force. After some time they are injected into the beam center where they are accelerated by the mechanisms described above. The injection mechanism is explained in the following section.
	
The influence of preplasma scale length on electron energies was examined by two additional simulations with p-polarized pulse. Shorter preplasma had a scale length $L_s=2\lambda~(1.6\rm{\mu m})$ and longer preplasma scale length was set to $L_s=6.25\lambda~(5.0\rm{\mu m})$. The trend was obvious - longer preplasma resulted in higher electron energies. Maximum $p_x$ for shortest preplasma was $\approx150m_ec$, maximum $p_x$ for moderate preplasma ($3.75\lambda,3.0\rm{\mu m}$) was $\approx200m_ec$ and longest preplasma target resulted in electrons with almost $300m_ec$. 

Higher electron energies in the case of moderate preplasma were the result of longer interaction length of electrons with the standing wave. The longer interaction length is a known property of stochastic heating \cite{stoch_paradkar}. Since in the case of moderate preplasma the maximum possible distance of interaction with the standing wave was already achieved (defined by the length of laser pulse), the result of increase in electron energy for the longest examined preplasma is different. When laser pulse propagated in preplasma, the modulation of laser pulse amplitude started to be developed, which created the lower amplitude region in the center. Electrons were captured inside this region in the center of propagating laser pulse and they were pre-accelerated even before the laser reflection. The pre-acceleration of electrons before the stochastic heating led to the higher electron energies. Similar phenomena was observed in Ref.\onlinecite{preplasma_had}. It caused the higher energies of accelerated electrons, even though the capturing mechanism is assigned to different effect in the reference. 
	
	Interesting property of $p_x$ evolution for the majority of hot electrons was the negative momentum values just before the acceleration into the target. This was however the same feature of all electrons being accelerated in four distinct interaction regimes described.

\section{Standing wave electron injection}\label{ch:sw}

	\begin{figure*}
		\centering
		\includegraphics[width=1\textwidth]{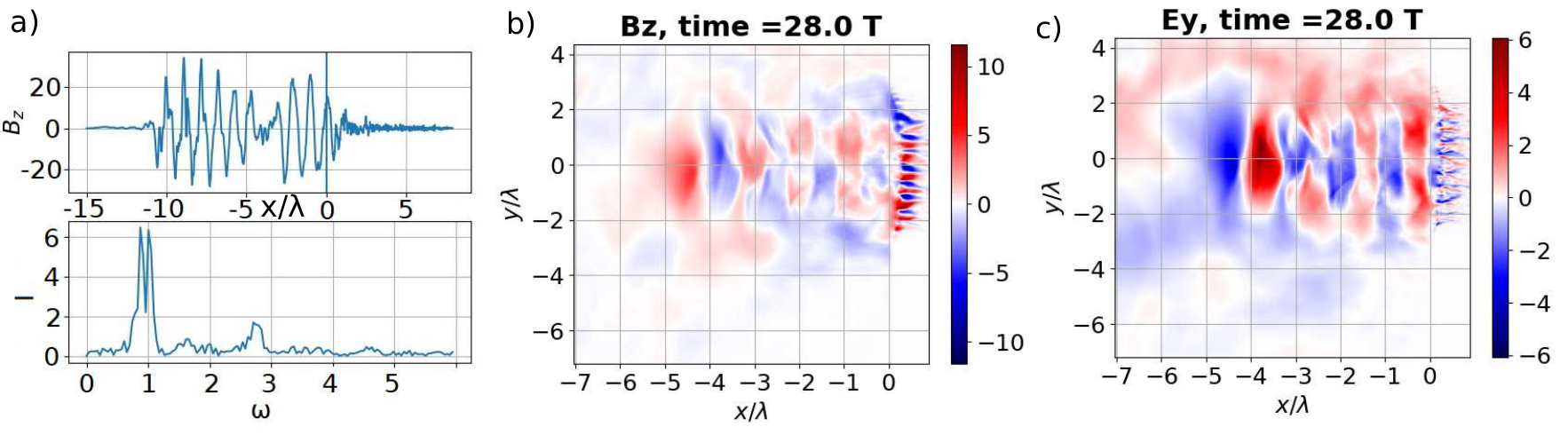}
		\caption{ (a) The magnetic field component of the standing wave for pulse with intensity $10^{21}~\rm{W/cm^2}$ at $y=1.5\lambda$ and its Fourier transform. $\omega=1$ corresponds to the incident wave and shifted peak of $\omega=0.9$ corresponds to the reflected wave. (b) Magnetic field in front of the overdense target averaged over single laser period in times between 27$T$ and 28$T$. (c) Electric field in front of the overdense target averaged over the same laser period.}
		\label{fig:pf-fields}
	\end{figure*}

	Ponderomotive force is well known for repelling particles from high field regions of laser pulses down the gradient. However, the nature of ponderomotive force in standing wave changes. It was previously shown \cite{kaplan_sw}, that ponderomotive force by the standing wave in relativistic regime can cause high-field regions to be attracting electrons. It was also shown that particles inside the standing wave exhibit complicated chaotic oscillations in Refs.~\onlinecite{bauer_pond,lehmann_attractors,sheng_stoch}. It can be seen in Fig. \ref{fig:pre-21-0-scatter} b) that only electrons at the position where standing wave was already present (marked by black circle) were injected to the beam axis. The most of electrons outside the circle is still on the sides of the beam, repelled by the incident pulse. 
	
	\begin{figure}
		\centering
		\includegraphics[width=.98\textwidth]{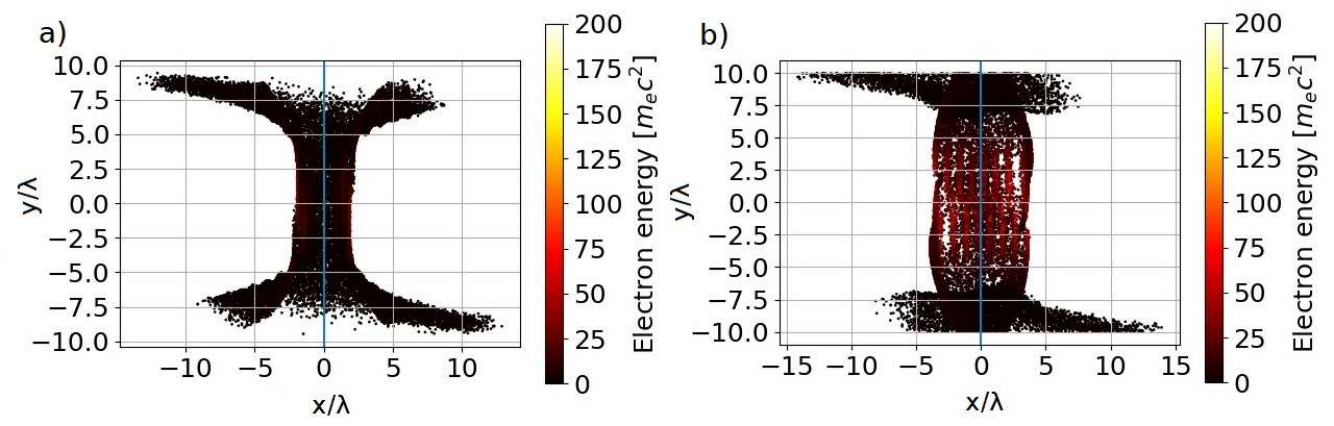}
		\caption{(a)Positions of electrons at time instant when laser pulses of intensity $10^{21}~\rm{W/cm^2}$ were propagating towards each other. Electrons are repelled by the ponderomotive force from the region where pulses are located. (b)Positions of electrons after laser pulses overlapped and created a standing wave present at $x=[-5,5]~\lambda$. Electrons are no longer repelled from the high-field regions and oscillate inside the standing wave. }
		\label{fig:pf-dens}
	\end{figure}

		To investigate closer the injection mechanism in standing wave into its high-field regions, we performed time-averaging of electromagnetic fields over one laser period shown in Fig. \ref{fig:pf-fields} b) and c). The region of non-zero averaged field can be observed in front of the target. Close to the beam axis, the field created due to the frequency shift of reflected pulse is present. Overdense plasma surface is pushed by the ponderomotive pressure which causes the red shift of reflected wave. This can be seen in the Fourier transform of standing wave. In Fig. \ref{fig:pf-fields}a), $\omega_0=1$ corresponds to the incident wave frequency and peak on the left with slightly lower frequency equal to $\omega_R=0.9$ is the frequency of reflected wave. This creates $\approx\lambda/2$ wide regions of non-zero averaged field with the periodic change of sign.
		
		 Further from the beam axis ($|y|>2\lambda$) the channel field \cite{khudik_channel,vranic_channel} dominates. It is created by the charge separation and acts on the repelled electrons in the direction towards the beam center. The channel field is also present close to the beam axis and the resulting time-averaged field is a superposition of the channel field and the field created by the red-shifted reflected pulse.

	\begin{figure*}
		\centering
		\includegraphics[width=1\textwidth]{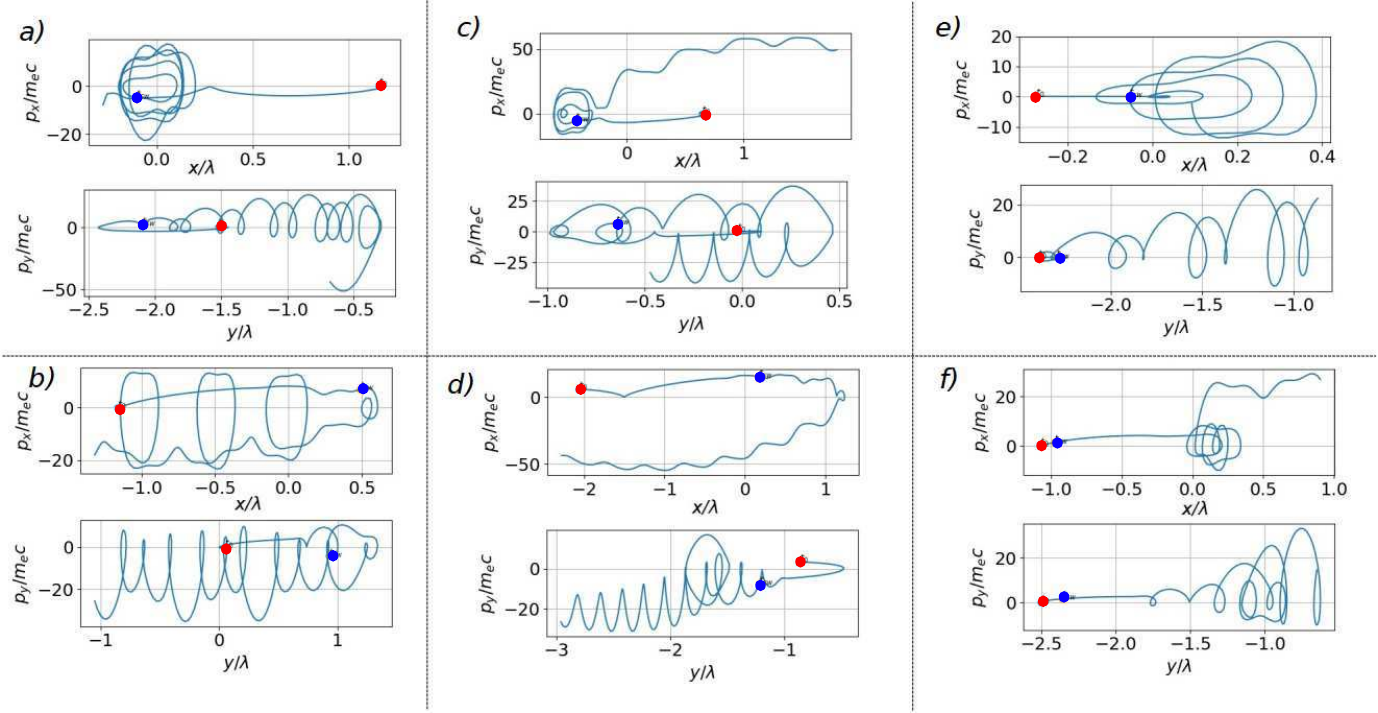}
		\caption{Example phase space trajectories of electrons in the field of standing EM wave. Red dots correspond to the initial time of simulation and blue dotd correspond to the time 15$T$ when pulses started to overlap. (a)-(d) were obtained for counter-propagating pulses with identical frequencies, (e)-(f) for the pulses with shifted frequencies. }
		\label{fig:pf-xpx}
	\end{figure*}	
	
	In order to examine the influence of standing wave on electron dynamics, we performed the simulation only with test particles interacting with two counter-propagating laser pulses with intensity $10^{21}~\rm{W/cm^2}$ and identical temporal and spatial profile like in the previous simulations. Such approach allows us to observe particle dynamics without effects of collective plasma behavior. Test particles were evenly spread across the simulation box and laser pulses crossed at $x=0\lambda$ at time $t=15\rm{T}$. As can be seen in Fig. \ref{fig:pf-dens} a), while pulses propagate towards each other, ponderomotive force pushes electrons in front of the pulse and on the sides in agreement with our PIC simulation of pulse interacting with preplasma. The vertical column of particles around the blue line represents the region between two counter-propagating pulses. Electrons in this region were either pushed by the pulses or they were initially placed there. Those electrons are also the source of electron population A in section \ref{ch:e21pre}. Fig. \ref{fig:pf-dens} b) shows positions of electrons after beams crossed. The wavefront of the laser pulse propagating from the right is located approximately at $x=-5\lambda$, and the wavefront of the pulse propagating from the left is located approximately at $x=5\lambda$. The region where particles are present coincides with the area where standing wave was already created due to the overlap of both pulses. This means that pushing electrons down the wave packet gradient is not the dominant ponderomotive force behavior and particles are allowed to oscillate inside the high-intensity region in the standing wave at such intensities. In later times when pulses propagate further, the region where particles are not repelled increases. This clearly demonstrates that the creation of standing wave by the reflected laser pulse had strong impact on the injection of electrons into the standing wave in the case of laser pulse with intensity $10^{21}~\rm{W/cm^2}$ incident on target with preplasma.

	Observation of electron trajectories in phase space shows that standing EM wave allows electrons to oscillate in the high-field regions instead of being reflected by the ponderomotive force. Trajectories vary greatly, however several patterns can be observed, see Fig. \ref{fig:pf-xpx}. Examples a)-d) correspond to electrons interacting with pulses with same frequency and e)-)f correspond to pulses with shifted frequencies. In Fig. \ref{fig:pf-xpx} a), $x-p_x$ trajectory oscillates around the position of $B_z$ field antinode and the node of $E_y$ field of standing wave. In $y-p_y$ trajectory, oscillation in positive $p_y$ values is visible while electron moves in the direction towards the beam axis at $y=0$. This is in agreement with trajectory obtained from PIC simulation in Fig. \ref{fig:pre-21-0-f}. In Fig. \ref{fig:pf-xpx} b), it can be seen how electron skips from one $B_z$ antinode to another in $x-p_x$ phase space. In $y-p_y$, electron oscillates in negative $p_y$ values and moves at first in direction towards the beam axis and after it continues in oscillation and heads towards lower pulse intensity regions. In Fig. \ref{fig:pf-xpx} c), electron at first oscillates around $B_z$ antinode in $x$-direction and afterwards is accelerated to momenta exceeding values of oscillation in the laser field. The $p_y$ momentum oscillates in positive values and after some time starts to oscillate in negative values. The moment when oscillation sign changes is identical to the time when electron starts to be accelerated in $x$-direction. In Fig. \ref{fig:pf-xpx} d), the immediate acceleration by the standing wave to high energies without oscillation around $E_y$ node is present. 
	
	The case in Fig. \ref{fig:pf-xpx} e) and f) demonstrates the impact of reflected wave's shifted frequency. The frequency of the pulse propagating from the left was set to $0.9\omega_0$, where $\omega_0$ is the frequency of the pulse propagating from the right boundary. The shift of oscillation center in $x$-direction is the only observed difference compared to the situation when pulses with identical frequencies couter-propagate.

The discussion above allows us to describe the electron injection in the following way. At first, electrons are repelled by the propagating laser pulse far from the beam axis. The channel field pushing electrons towards the beam center is simultaneously created. However, ponderomotive force overcomes the effect of channel field and keeps electrons on the sides. After the reflection of incident laser wave the standing wave is created and electrons are no longer repelled by the ponderomotive force. Instead, they are oscillating around the magnetic field antinodes. This allows electrons to be injected by the channel field closer to the beam axis where the motion is dominated by the standing wave. There, electrons are either accelerated in $x$-direction or they oscillate around the $B$ field antinode while they are injected further in transverse direction as shown in Fig. \ref{fig:18-0-xpx}.

	\section{Conclusion}
	
	We have demonstrated, that hot electron acceleration mechanisms strongly differ depending on the laser intensity and density profile. For the case of intensity $10^{18}~\rm{W/cm^2}$ and steep plasma gradient, we showed that electron is injected into the vacuum by the electrostatic field where it interacts with the standing wave and starts to be accelerated back into the plasma at times, when magnetic field changes its polarity twice a laser period. The increase of intensity to $10^{21}~\rm{W/cm^2}$ caused the deformation of overdense plasma boundary, which resulted in chaotic oscillations of electrons before the acceleration. Without the presence of preplasma, electrons were accelerated at vacuum-plasma boundary because of $\vec{j}\times\vec{B}$ heating. 
	
	The presence of preplasma caused, that all electrons were accelerated in the underdense region, which resulted in significantly higher hot electron temperatures. Significant impact of pulse polarization on electron temperature was explained by the background field of plasma channel. For intensity $10^{18}~\rm{W/cm^2}$, irreversible energy gain was not present until the moment, when reflected wave reached the position of an electron. Afterwards, electrons were accelerated towards the overdense part of the target while the stochastic acceleration along with longitudinal electric field present in preplasma were dominant in both regimes. When laser pulse with intensity $10^{21}~\rm{W/cm^2}$ was incident on the target, the first population of hot electrons was pre-accelerated by the front of laser pulse. The second population of hot electrons was initially repelled from the high-field regions by the ponderomotive force of incident wave. However, when standing wave was created, they were injected back into the beam axis region along the magnetic field antinode and accelerated afterwards. This injection is a result of relativistic electron dynamics in the field of standing electromagnetic wave with the combination of plasma channel field. It was demonstrated, that longer preplasma scale length results in higher electron energies. 
	
	\begin{acknowledgments}
	This work was partially supported by CTU project “Research on optical (nano)structures and laser plasma” No. SGS19/192/OHK4/3T/14.
		
	Computational resources were supplied by the project "e-Infrastruktura CZ" (e-INFRA LM2018140) provided within the program Projects of Large Research, Development and Innovations Infrastructures.
		
	J. P. also acknowledges the support and computational resources obtained from European Regional Development Fund - Project CAAS ("Center of Advanced Applied Sciences") No. CZ.02.1.01/0.0/0.0/16$\_$019/0000778.
		
	Fruitful discussions with M. Vranic (IPFN, Instituto Superior Técnico) and D. Maslarova (IPP, The Czech Academy of Sciences) are gratefully acknowledged.
	\end{acknowledgments}
	
		\section*{Data Availability Statement}
	The data that support the findings of this study are available from the corresponding author upon reasonable request.

	\bibliography{bibliography}

\begin{thebibliography}{40}%
\makeatletter
\providecommand \@ifxundefined [1]{%
 \@ifx{#1\undefined}
}%
\providecommand \@ifnum [1]{%
 \ifnum #1\expandafter \@firstoftwo
 \else \expandafter \@secondoftwo
 \fi
}%
\providecommand \@ifx [1]{%
 \ifx #1\expandafter \@firstoftwo
 \else \expandafter \@secondoftwo
 \fi
}%
\providecommand \natexlab [1]{#1}%
\providecommand \enquote  [1]{``#1''}%
\providecommand \bibnamefont  [1]{#1}%
\providecommand \bibfnamefont [1]{#1}%
\providecommand \citenamefont [1]{#1}%
\providecommand \href@noop [0]{\@secondoftwo}%
\providecommand \href [0]{\begingroup \@sanitize@url \@href}%
\providecommand \@href[1]{\@@startlink{#1}\@@href}%
\providecommand \@@href[1]{\endgroup#1\@@endlink}%
\providecommand \@sanitize@url [0]{\catcode `\\12\catcode `\$12\catcode
  `\&12\catcode `\#12\catcode `\^12\catcode `\_12\catcode `\%12\relax}%
\providecommand \@@startlink[1]{}%
\providecommand \@@endlink[0]{}%
\providecommand \url  [0]{\begingroup\@sanitize@url \@url }%
\providecommand \@url [1]{\endgroup\@href {#1}{\urlprefix }}%
\providecommand \urlprefix  [0]{URL }%
\providecommand \Eprint [0]{\href }%
\providecommand \doibase [0]{http://dx.doi.org/}%
\providecommand \selectlanguage [0]{\@gobble}%
\providecommand \bibinfo  [0]{\@secondoftwo}%
\providecommand \bibfield  [0]{\@secondoftwo}%
\providecommand \translation [1]{[#1]}%
\providecommand \BibitemOpen [0]{}%
\providecommand \bibitemStop [0]{}%
\providecommand \bibitemNoStop [0]{.\EOS\space}%
\providecommand \EOS [0]{\spacefactor3000\relax}%
\providecommand \BibitemShut  [1]{\csname bibitem#1\endcsname}%
\let\auto@bib@innerbib\@empty
\bibitem [{\citenamefont {Wilks}\ \emph {et~al.}(2001)\citenamefont {Wilks},
  \citenamefont {Langdon}, \citenamefont {Cowan}, \citenamefont {Roth},
  \citenamefont {Singh}, \citenamefont {Hatchett}, \citenamefont {Key},
  \citenamefont {Pennington}, \citenamefont {MacKinnon},\ and\ \citenamefont
  {Snavely}}]{wilks_tnsa}%
  \BibitemOpen
  \bibfield  {author} {\bibinfo {author} {\bibfnamefont {S.~C.}\ \bibnamefont
  {Wilks}}, \bibinfo {author} {\bibfnamefont {A.~B.}\ \bibnamefont {Langdon}},
  \bibinfo {author} {\bibfnamefont {T.~E.}\ \bibnamefont {Cowan}}, \bibinfo
  {author} {\bibfnamefont {M.}~\bibnamefont {Roth}}, \bibinfo {author}
  {\bibfnamefont {M.}~\bibnamefont {Singh}}, \bibinfo {author} {\bibfnamefont
  {S.}~\bibnamefont {Hatchett}}, \bibinfo {author} {\bibfnamefont {M.~H.}\
  \bibnamefont {Key}}, \bibinfo {author} {\bibfnamefont {D.}~\bibnamefont
  {Pennington}}, \bibinfo {author} {\bibfnamefont {A.}~\bibnamefont
  {MacKinnon}}, \ and\ \bibinfo {author} {\bibfnamefont {R.~A.}\ \bibnamefont
  {Snavely}},\ }\bibfield  {title} {\enquote {\bibinfo {title} {Energetic
  proton generation in ultra-intense laser–solid interactions},}\ }\href
  {\doibase 10.1063/1.1333697} {\bibfield  {journal} {\bibinfo  {journal}
  {Physics of Plasmas}\ }\textbf {\bibinfo {volume} {8}},\ \bibinfo {pages}
  {542--549} (\bibinfo {year} {2001})},\ \Eprint
  {http://arxiv.org/abs/https://doi.org/10.1063/1.1333697}
  {https://doi.org/10.1063/1.1333697} \BibitemShut {NoStop}%
\bibitem [{\citenamefont {Norreys}\ \emph {et~al.}(1996)\citenamefont
  {Norreys}, \citenamefont {Zepf}, \citenamefont {Moustaizis}, \citenamefont
  {Fews}, \citenamefont {Zhang}, \citenamefont {Lee}, \citenamefont
  {Bakarezos}, \citenamefont {Danson}, \citenamefont {Dyson}, \citenamefont
  {Gibbon}, \citenamefont {Loukakos}, \citenamefont {Neely}, \citenamefont
  {Walsh}, \citenamefont {Wark},\ and\ \citenamefont {Dangor}}]{norreys_hhg}%
  \BibitemOpen
  \bibfield  {author} {\bibinfo {author} {\bibfnamefont {P.~A.}\ \bibnamefont
  {Norreys}}, \bibinfo {author} {\bibfnamefont {M.}~\bibnamefont {Zepf}},
  \bibinfo {author} {\bibfnamefont {S.}~\bibnamefont {Moustaizis}}, \bibinfo
  {author} {\bibfnamefont {A.~P.}\ \bibnamefont {Fews}}, \bibinfo {author}
  {\bibfnamefont {J.}~\bibnamefont {Zhang}}, \bibinfo {author} {\bibfnamefont
  {P.}~\bibnamefont {Lee}}, \bibinfo {author} {\bibfnamefont {M.}~\bibnamefont
  {Bakarezos}}, \bibinfo {author} {\bibfnamefont {C.~N.}\ \bibnamefont
  {Danson}}, \bibinfo {author} {\bibfnamefont {A.}~\bibnamefont {Dyson}},
  \bibinfo {author} {\bibfnamefont {P.}~\bibnamefont {Gibbon}}, \bibinfo
  {author} {\bibfnamefont {P.}~\bibnamefont {Loukakos}}, \bibinfo {author}
  {\bibfnamefont {D.}~\bibnamefont {Neely}}, \bibinfo {author} {\bibfnamefont
  {F.~N.}\ \bibnamefont {Walsh}}, \bibinfo {author} {\bibfnamefont {J.~S.}\
  \bibnamefont {Wark}}, \ and\ \bibinfo {author} {\bibfnamefont {A.~E.}\
  \bibnamefont {Dangor}},\ }\bibfield  {title} {\enquote {\bibinfo {title}
  {Efficient extreme uv harmonics generated from picosecond laser pulse
  interactions with solid targets},}\ }\href {\doibase
  10.1103/PhysRevLett.76.1832} {\bibfield  {journal} {\bibinfo  {journal}
  {Phys. Rev. Lett.}\ }\textbf {\bibinfo {volume} {76}},\ \bibinfo {pages}
  {1832--1835} (\bibinfo {year} {1996})}\BibitemShut {NoStop}%
\bibitem [{\citenamefont {Craxton}\ \emph {et~al.}(2015)\citenamefont
  {Craxton}, \citenamefont {Anderson}, \citenamefont {Boehly}, \citenamefont
  {Goncharov}, \citenamefont {Harding}, \citenamefont {Knauer}, \citenamefont
  {McCrory}, \citenamefont {McKenty}, \citenamefont {Meyerhofer}, \citenamefont
  {Myatt}, \citenamefont {Schmitt}, \citenamefont {Sethian}, \citenamefont
  {Short}, \citenamefont {Skupsky}, \citenamefont {Theobald}, \citenamefont
  {Kruer}, \citenamefont {Tanaka}, \citenamefont {Betti}, \citenamefont
  {Collins}, \citenamefont {Delettrez}, \citenamefont {Hu}, \citenamefont
  {Marozas}, \citenamefont {Maximov}, \citenamefont {Michel}, \citenamefont
  {Radha}, \citenamefont {Regan}, \citenamefont {Sangster}, \citenamefont
  {Seka}, \citenamefont {Solodov}, \citenamefont {Soures}, \citenamefont
  {Stoeckl},\ and\ \citenamefont {Zuegel}}]{craxton_icf}%
  \BibitemOpen
  \bibfield  {author} {\bibinfo {author} {\bibfnamefont {R.~S.}\ \bibnamefont
  {Craxton}}, \bibinfo {author} {\bibfnamefont {K.~S.}\ \bibnamefont
  {Anderson}}, \bibinfo {author} {\bibfnamefont {T.~R.}\ \bibnamefont
  {Boehly}}, \bibinfo {author} {\bibfnamefont {V.~N.}\ \bibnamefont
  {Goncharov}}, \bibinfo {author} {\bibfnamefont {D.~R.}\ \bibnamefont
  {Harding}}, \bibinfo {author} {\bibfnamefont {J.~P.}\ \bibnamefont {Knauer}},
  \bibinfo {author} {\bibfnamefont {R.~L.}\ \bibnamefont {McCrory}}, \bibinfo
  {author} {\bibfnamefont {P.~W.}\ \bibnamefont {McKenty}}, \bibinfo {author}
  {\bibfnamefont {D.~D.}\ \bibnamefont {Meyerhofer}}, \bibinfo {author}
  {\bibfnamefont {J.~F.}\ \bibnamefont {Myatt}}, \bibinfo {author}
  {\bibfnamefont {A.~J.}\ \bibnamefont {Schmitt}}, \bibinfo {author}
  {\bibfnamefont {J.~D.}\ \bibnamefont {Sethian}}, \bibinfo {author}
  {\bibfnamefont {R.~W.}\ \bibnamefont {Short}}, \bibinfo {author}
  {\bibfnamefont {S.}~\bibnamefont {Skupsky}}, \bibinfo {author} {\bibfnamefont
  {W.}~\bibnamefont {Theobald}}, \bibinfo {author} {\bibfnamefont {W.~L.}\
  \bibnamefont {Kruer}}, \bibinfo {author} {\bibfnamefont {K.}~\bibnamefont
  {Tanaka}}, \bibinfo {author} {\bibfnamefont {R.}~\bibnamefont {Betti}},
  \bibinfo {author} {\bibfnamefont {T.~J.~B.}\ \bibnamefont {Collins}},
  \bibinfo {author} {\bibfnamefont {J.~A.}\ \bibnamefont {Delettrez}}, \bibinfo
  {author} {\bibfnamefont {S.~X.}\ \bibnamefont {Hu}}, \bibinfo {author}
  {\bibfnamefont {J.~A.}\ \bibnamefont {Marozas}}, \bibinfo {author}
  {\bibfnamefont {A.~V.}\ \bibnamefont {Maximov}}, \bibinfo {author}
  {\bibfnamefont {D.~T.}\ \bibnamefont {Michel}}, \bibinfo {author}
  {\bibfnamefont {P.~B.}\ \bibnamefont {Radha}}, \bibinfo {author}
  {\bibfnamefont {S.~P.}\ \bibnamefont {Regan}}, \bibinfo {author}
  {\bibfnamefont {T.~C.}\ \bibnamefont {Sangster}}, \bibinfo {author}
  {\bibfnamefont {W.}~\bibnamefont {Seka}}, \bibinfo {author} {\bibfnamefont
  {A.~A.}\ \bibnamefont {Solodov}}, \bibinfo {author} {\bibfnamefont {J.~M.}\
  \bibnamefont {Soures}}, \bibinfo {author} {\bibfnamefont {C.}~\bibnamefont
  {Stoeckl}}, \ and\ \bibinfo {author} {\bibfnamefont {J.~D.}\ \bibnamefont
  {Zuegel}},\ }\bibfield  {title} {\enquote {\bibinfo {title} {Direct-drive
  inertial confinement fusion: A review},}\ }\href {\doibase 10.1063/1.4934714}
  {\bibfield  {journal} {\bibinfo  {journal} {Physics of Plasmas}\ }\textbf
  {\bibinfo {volume} {22}},\ \bibinfo {pages} {110501} (\bibinfo {year}
  {2015})},\ \Eprint {http://arxiv.org/abs/https://doi.org/10.1063/1.4934714}
  {https://doi.org/10.1063/1.4934714} \BibitemShut {NoStop}%
\bibitem [{\citenamefont {Bulanov}\ \emph {et~al.}(2015)\citenamefont
  {Bulanov}, \citenamefont {Esirkepov}, \citenamefont {Kando}, \citenamefont
  {Koga}, \citenamefont {Kondo},\ and\ \citenamefont {Korn}}]{bulanov_astro}%
  \BibitemOpen
  \bibfield  {author} {\bibinfo {author} {\bibfnamefont {S.}~\bibnamefont
  {Bulanov}}, \bibinfo {author} {\bibfnamefont {T.}~\bibnamefont {Esirkepov}},
  \bibinfo {author} {\bibfnamefont {M.}~\bibnamefont {Kando}}, \bibinfo
  {author} {\bibfnamefont {J.}~\bibnamefont {Koga}}, \bibinfo {author}
  {\bibfnamefont {K.}~\bibnamefont {Kondo}}, \ and\ \bibinfo {author}
  {\bibfnamefont {G.}~\bibnamefont {Korn}},\ }\bibfield  {title} {\enquote
  {\bibinfo {title} {On the problems of relativistic laboratory astrophysics
  and fundamental physics with super powerful lasers},}\ }\href {\doibase
  10.1134/S1063780X15010018} {\bibfield  {journal} {\bibinfo  {journal} {Plasma
  Physics Reports}\ }\textbf {\bibinfo {volume} {41}},\ \bibinfo {pages}
  {1--51} (\bibinfo {year} {2015})}\BibitemShut {NoStop}%
\bibitem [{\citenamefont {Macchi}, \citenamefont {Borghesi},\ and\
  \citenamefont {Passoni}(2013)}]{macchi_review}%
  \BibitemOpen
  \bibfield  {author} {\bibinfo {author} {\bibfnamefont {A.}~\bibnamefont
  {Macchi}}, \bibinfo {author} {\bibfnamefont {M.}~\bibnamefont {Borghesi}}, \
  and\ \bibinfo {author} {\bibfnamefont {M.}~\bibnamefont {Passoni}},\
  }\bibfield  {title} {\enquote {\bibinfo {title} {Ion acceleration by
  superintense laser-plasma interaction},}\ }\href {\doibase
  10.1103/RevModPhys.85.751} {\bibfield  {journal} {\bibinfo  {journal} {Rev.
  Mod. Phys.}\ }\textbf {\bibinfo {volume} {85}},\ \bibinfo {pages} {751--793}
  (\bibinfo {year} {2013})}\BibitemShut {NoStop}%
\bibitem [{\citenamefont {Beg}\ \emph {et~al.}(1997)\citenamefont {Beg},
  \citenamefont {Bell}, \citenamefont {Dangor}, \citenamefont {Danson},
  \citenamefont {Fews}, \citenamefont {Glinsky}, \citenamefont {Hammel},
  \citenamefont {Lee}, \citenamefont {Norreys},\ and\ \citenamefont
  {Tatarakis}}]{beg_scaling}%
  \BibitemOpen
  \bibfield  {author} {\bibinfo {author} {\bibfnamefont {F.~N.}\ \bibnamefont
  {Beg}}, \bibinfo {author} {\bibfnamefont {A.~R.}\ \bibnamefont {Bell}},
  \bibinfo {author} {\bibfnamefont {A.~E.}\ \bibnamefont {Dangor}}, \bibinfo
  {author} {\bibfnamefont {C.~N.}\ \bibnamefont {Danson}}, \bibinfo {author}
  {\bibfnamefont {A.~P.}\ \bibnamefont {Fews}}, \bibinfo {author}
  {\bibfnamefont {M.~E.}\ \bibnamefont {Glinsky}}, \bibinfo {author}
  {\bibfnamefont {B.~A.}\ \bibnamefont {Hammel}}, \bibinfo {author}
  {\bibfnamefont {P.}~\bibnamefont {Lee}}, \bibinfo {author} {\bibfnamefont
  {P.~A.}\ \bibnamefont {Norreys}}, \ and\ \bibinfo {author} {\bibfnamefont
  {M.}~\bibnamefont {Tatarakis}},\ }\bibfield  {title} {\enquote {\bibinfo
  {title} {A study of picosecond laser–solid interactions up to
  $10^{19}~\rm{Wcm}^{-2}$},}\ }\href {\doibase 10.1063/1.872103} {\bibfield
  {journal} {\bibinfo  {journal} {Physics of Plasmas}\ }\textbf {\bibinfo
  {volume} {4}},\ \bibinfo {pages} {447--457} (\bibinfo {year} {1997})},\
  \Eprint {http://arxiv.org/abs/https://doi.org/10.1063/1.872103}
  {https://doi.org/10.1063/1.872103} \BibitemShut {NoStop}%
\bibitem [{\citenamefont {Kluge}\ \emph {et~al.}(2011)\citenamefont {Kluge},
  \citenamefont {Cowan}, \citenamefont {Debus}, \citenamefont {Schramm},
  \citenamefont {Zeil},\ and\ \citenamefont {Bussmann}}]{kluge_scaling}%
  \BibitemOpen
  \bibfield  {author} {\bibinfo {author} {\bibfnamefont {T.}~\bibnamefont
  {Kluge}}, \bibinfo {author} {\bibfnamefont {T.}~\bibnamefont {Cowan}},
  \bibinfo {author} {\bibfnamefont {A.}~\bibnamefont {Debus}}, \bibinfo
  {author} {\bibfnamefont {U.}~\bibnamefont {Schramm}}, \bibinfo {author}
  {\bibfnamefont {K.}~\bibnamefont {Zeil}}, \ and\ \bibinfo {author}
  {\bibfnamefont {M.}~\bibnamefont {Bussmann}},\ }\bibfield  {title} {\enquote
  {\bibinfo {title} {Electron temperature scaling in laser interaction with
  solids},}\ }\href {\doibase 10.1103/PhysRevLett.107.205003} {\bibfield
  {journal} {\bibinfo  {journal} {Phys. Rev. Lett.}\ }\textbf {\bibinfo
  {volume} {107}},\ \bibinfo {pages} {205003} (\bibinfo {year}
  {2011})}\BibitemShut {NoStop}%
\bibitem [{\citenamefont {Wilks}\ \emph {et~al.}(1992)\citenamefont {Wilks},
  \citenamefont {Kruer}, \citenamefont {Tabak},\ and\ \citenamefont
  {Langdon}}]{wilks_abs}%
  \BibitemOpen
  \bibfield  {author} {\bibinfo {author} {\bibfnamefont {S.~C.}\ \bibnamefont
  {Wilks}}, \bibinfo {author} {\bibfnamefont {W.~L.}\ \bibnamefont {Kruer}},
  \bibinfo {author} {\bibfnamefont {M.}~\bibnamefont {Tabak}}, \ and\ \bibinfo
  {author} {\bibfnamefont {A.~B.}\ \bibnamefont {Langdon}},\ }\bibfield
  {title} {\enquote {\bibinfo {title} {Absorption of ultra-intense laser
  pulses},}\ }\href {\doibase 10.1103/PhysRevLett.69.1383} {\bibfield
  {journal} {\bibinfo  {journal} {Phys. Rev. Lett.}\ }\textbf {\bibinfo
  {volume} {69}},\ \bibinfo {pages} {1383--1386} (\bibinfo {year}
  {1992})}\BibitemShut {NoStop}%
\bibitem [{\citenamefont {Denavit}(1992)}]{denavit_abs}%
  \BibitemOpen
  \bibfield  {author} {\bibinfo {author} {\bibfnamefont {J.}~\bibnamefont
  {Denavit}},\ }\bibfield  {title} {\enquote {\bibinfo {title} {Absorption of
  high-intensity subpicosecond lasers on solid density targets},}\ }\href
  {\doibase 10.1103/PhysRevLett.69.3052} {\bibfield  {journal} {\bibinfo
  {journal} {Phys. Rev. Lett.}\ }\textbf {\bibinfo {volume} {69}},\ \bibinfo
  {pages} {3052--3055} (\bibinfo {year} {1992})}\BibitemShut {NoStop}%
\bibitem [{\citenamefont {Chen}\ \emph {et~al.}(2009)\citenamefont {Chen},
  \citenamefont {Wilks}, \citenamefont {Kruer}, \citenamefont {Patel},\ and\
  \citenamefont {Shepherd}}]{chen_electrons}%
  \BibitemOpen
  \bibfield  {author} {\bibinfo {author} {\bibfnamefont {H.}~\bibnamefont
  {Chen}}, \bibinfo {author} {\bibfnamefont {S.~C.}\ \bibnamefont {Wilks}},
  \bibinfo {author} {\bibfnamefont {W.~L.}\ \bibnamefont {Kruer}}, \bibinfo
  {author} {\bibfnamefont {P.~K.}\ \bibnamefont {Patel}}, \ and\ \bibinfo
  {author} {\bibfnamefont {R.}~\bibnamefont {Shepherd}},\ }\bibfield  {title}
  {\enquote {\bibinfo {title} {Hot electron energy distributions from
  ultraintense laser solid interactions},}\ }\href {\doibase 10.1063/1.3080197}
  {\bibfield  {journal} {\bibinfo  {journal} {Physics of Plasmas}\ }\textbf
  {\bibinfo {volume} {16}},\ \bibinfo {pages} {020705} (\bibinfo {year}
  {2009})},\ \Eprint {http://arxiv.org/abs/https://doi.org/10.1063/1.3080197}
  {https://doi.org/10.1063/1.3080197} \BibitemShut {NoStop}%
\bibitem [{\citenamefont {Culfa}\ \emph {et~al.}(2014)\citenamefont {Culfa},
  \citenamefont {Tallents}, \citenamefont {Wagenaars}, \citenamefont {Ridgers},
  \citenamefont {Dance}, \citenamefont {Rossall}, \citenamefont {Gray},
  \citenamefont {McKenna}, \citenamefont {Brown}, \citenamefont {James},
  \citenamefont {Hoarty}, \citenamefont {Booth}, \citenamefont {Robinson},
  \citenamefont {Lancaster}, \citenamefont {Pikuz}, \citenamefont {Faenov},
  \citenamefont {Kampfer}, \citenamefont {Schulze}, \citenamefont {Uschmann},\
  and\ \citenamefont {Woolsey}}]{culfa_electrons}%
  \BibitemOpen
  \bibfield  {author} {\bibinfo {author} {\bibfnamefont {O.}~\bibnamefont
  {Culfa}}, \bibinfo {author} {\bibfnamefont {G.~J.}\ \bibnamefont {Tallents}},
  \bibinfo {author} {\bibfnamefont {E.}~\bibnamefont {Wagenaars}}, \bibinfo
  {author} {\bibfnamefont {C.~P.}\ \bibnamefont {Ridgers}}, \bibinfo {author}
  {\bibfnamefont {R.~J.}\ \bibnamefont {Dance}}, \bibinfo {author}
  {\bibfnamefont {A.~K.}\ \bibnamefont {Rossall}}, \bibinfo {author}
  {\bibfnamefont {R.~J.}\ \bibnamefont {Gray}}, \bibinfo {author}
  {\bibfnamefont {P.}~\bibnamefont {McKenna}}, \bibinfo {author} {\bibfnamefont
  {C.~D.~R.}\ \bibnamefont {Brown}}, \bibinfo {author} {\bibfnamefont {S.~F.}\
  \bibnamefont {James}}, \bibinfo {author} {\bibfnamefont {D.~J.}\ \bibnamefont
  {Hoarty}}, \bibinfo {author} {\bibfnamefont {N.}~\bibnamefont {Booth}},
  \bibinfo {author} {\bibfnamefont {A.~P.~L.}\ \bibnamefont {Robinson}},
  \bibinfo {author} {\bibfnamefont {K.~L.}\ \bibnamefont {Lancaster}}, \bibinfo
  {author} {\bibfnamefont {S.~A.}\ \bibnamefont {Pikuz}}, \bibinfo {author}
  {\bibfnamefont {A.~Y.}\ \bibnamefont {Faenov}}, \bibinfo {author}
  {\bibfnamefont {T.}~\bibnamefont {Kampfer}}, \bibinfo {author} {\bibfnamefont
  {K.~S.}\ \bibnamefont {Schulze}}, \bibinfo {author} {\bibfnamefont
  {I.}~\bibnamefont {Uschmann}}, \ and\ \bibinfo {author} {\bibfnamefont
  {N.~C.}\ \bibnamefont {Woolsey}},\ }\bibfield  {title} {\enquote {\bibinfo
  {title} {Hot electron production in laser solid interactions with a
  controlled pre-pulse},}\ }\href {\doibase 10.1063/1.4870633} {\bibfield
  {journal} {\bibinfo  {journal} {Physics of Plasmas}\ }\textbf {\bibinfo
  {volume} {21}},\ \bibinfo {pages} {043106} (\bibinfo {year} {2014})},\
  \Eprint {http://arxiv.org/abs/https://doi.org/10.1063/1.4870633}
  {https://doi.org/10.1063/1.4870633} \BibitemShut {NoStop}%
\bibitem [{\citenamefont {Debayle}\ \emph {et~al.}(2013)\citenamefont
  {Debayle}, \citenamefont {Sanz}, \citenamefont {Gremillet},\ and\
  \citenamefont {Mima}}]{debayle}%
  \BibitemOpen
  \bibfield  {author} {\bibinfo {author} {\bibfnamefont {A.}~\bibnamefont
  {Debayle}}, \bibinfo {author} {\bibfnamefont {J.}~\bibnamefont {Sanz}},
  \bibinfo {author} {\bibfnamefont {L.}~\bibnamefont {Gremillet}}, \ and\
  \bibinfo {author} {\bibfnamefont {K.}~\bibnamefont {Mima}},\ }\bibfield
  {title} {\enquote {\bibinfo {title} {Toward a self-consistent model of the
  interaction between an ultra-intense, normally incident laser pulse with an
  overdense plasma},}\ }\href {\doibase 10.1063/1.4807335} {\bibfield
  {journal} {\bibinfo  {journal} {Physics of Plasmas}\ }\textbf {\bibinfo
  {volume} {20}},\ \bibinfo {pages} {053107} (\bibinfo {year} {2013})},\
  \Eprint {http://arxiv.org/abs/https://doi.org/10.1063/1.4807335}
  {https://doi.org/10.1063/1.4807335} \BibitemShut {NoStop}%
\bibitem [{\citenamefont {Bulanov}\ \emph {et~al.}(2013)\citenamefont
  {Bulanov}, \citenamefont {Esirkepov}, \citenamefont {Kando}, \citenamefont
  {Bulanov}, \citenamefont {Rykovanov},\ and\ \citenamefont
  {Pegoraro}}]{bulanov_foil}%
  \BibitemOpen
  \bibfield  {author} {\bibinfo {author} {\bibfnamefont {S.~V.}\ \bibnamefont
  {Bulanov}}, \bibinfo {author} {\bibfnamefont {T.~Z.}\ \bibnamefont
  {Esirkepov}}, \bibinfo {author} {\bibfnamefont {M.}~\bibnamefont {Kando}},
  \bibinfo {author} {\bibfnamefont {S.~S.}\ \bibnamefont {Bulanov}}, \bibinfo
  {author} {\bibfnamefont {S.~G.}\ \bibnamefont {Rykovanov}}, \ and\ \bibinfo
  {author} {\bibfnamefont {F.}~\bibnamefont {Pegoraro}},\ }\bibfield  {title}
  {\enquote {\bibinfo {title} {Strong field electrodynamics of a thin foil},}\
  }\href {\doibase 10.1063/1.4848758} {\bibfield  {journal} {\bibinfo
  {journal} {Physics of Plasmas}\ }\textbf {\bibinfo {volume} {20}},\ \bibinfo
  {pages} {123114} (\bibinfo {year} {2013})},\ \Eprint
  {http://arxiv.org/abs/https://doi.org/10.1063/1.4848758}
  {https://doi.org/10.1063/1.4848758} \BibitemShut {NoStop}%
\bibitem [{\citenamefont {{Kemp}}\ \emph {et~al.}(2013)\citenamefont {{Kemp}},
  \citenamefont {{Link}}, \citenamefont {{Ping}}, \citenamefont {{Schumacher}},
  \citenamefont {{Freeman}},\ and\ \citenamefont {{Patel}}}]{kemp}%
  \BibitemOpen
  \bibfield  {author} {\bibinfo {author} {\bibfnamefont {G.~E.}\ \bibnamefont
  {{Kemp}}}, \bibinfo {author} {\bibfnamefont {A.}~\bibnamefont {{Link}}},
  \bibinfo {author} {\bibfnamefont {Y.}~\bibnamefont {{Ping}}}, \bibinfo
  {author} {\bibfnamefont {D.~W.}\ \bibnamefont {{Schumacher}}}, \bibinfo
  {author} {\bibfnamefont {R.~R.}\ \bibnamefont {{Freeman}}}, \ and\ \bibinfo
  {author} {\bibfnamefont {P.~K.}\ \bibnamefont {{Patel}}},\ }\bibfield
  {title} {\enquote {\bibinfo {title} {{Coupling of laser energy into
  hot-electrons in high-contrast relativistic laser-plasma interactions}},}\
  }\href {\doibase 10.1063/1.4794961} {\bibfield  {journal} {\bibinfo
  {journal} {Physics of Plasmas}\ }\textbf {\bibinfo {volume} {20}},\ \bibinfo
  {eid} {033104} (\bibinfo {year} {2013})}\BibitemShut {NoStop}%
\bibitem [{\citenamefont {May}\ \emph {et~al.}(2011)\citenamefont {May},
  \citenamefont {Tonge}, \citenamefont {Fiuza}, \citenamefont {Fonseca},
  \citenamefont {Silva}, \citenamefont {Ren},\ and\ \citenamefont
  {Mori}}]{2a0}%
  \BibitemOpen
  \bibfield  {author} {\bibinfo {author} {\bibfnamefont {J.}~\bibnamefont
  {May}}, \bibinfo {author} {\bibfnamefont {J.}~\bibnamefont {Tonge}}, \bibinfo
  {author} {\bibfnamefont {F.}~\bibnamefont {Fiuza}}, \bibinfo {author}
  {\bibfnamefont {R.~A.}\ \bibnamefont {Fonseca}}, \bibinfo {author}
  {\bibfnamefont {L.~O.}\ \bibnamefont {Silva}}, \bibinfo {author}
  {\bibfnamefont {C.}~\bibnamefont {Ren}}, \ and\ \bibinfo {author}
  {\bibfnamefont {W.~B.}\ \bibnamefont {Mori}},\ }\bibfield  {title} {\enquote
  {\bibinfo {title} {Mechanism of generating fast electrons by an intense laser
  at a steep overdense interface},}\ }\href {\doibase
  10.1103/PhysRevE.84.025401} {\bibfield  {journal} {\bibinfo  {journal} {Phys.
  Rev. E}\ }\textbf {\bibinfo {volume} {84}},\ \bibinfo {pages} {025401}
  (\bibinfo {year} {2011})}\BibitemShut {NoStop}%
\bibitem [{\citenamefont {Hüller}\ \emph {et~al.}(2019)\citenamefont
  {Hüller}, \citenamefont {Porzio}, \citenamefont {Adam},\ and\ \citenamefont
  {Héron}}]{huller_electrons}%
  \BibitemOpen
  \bibfield  {author} {\bibinfo {author} {\bibfnamefont {S.}~\bibnamefont
  {Hüller}}, \bibinfo {author} {\bibfnamefont {A.}~\bibnamefont {Porzio}},
  \bibinfo {author} {\bibfnamefont {J.-C.}\ \bibnamefont {Adam}}, \ and\
  \bibinfo {author} {\bibfnamefont {A.}~\bibnamefont {Héron}},\ }\bibfield
  {title} {\enquote {\bibinfo {title} {On the non-thermal nature of
  distributions of electrons accelerated by high intensity lasers at the
  vacuum-plasma interface},}\ }\href {\doibase 10.1063/1.5111934} {\bibfield
  {journal} {\bibinfo  {journal} {Physics of Plasmas}\ }\textbf {\bibinfo
  {volume} {26}},\ \bibinfo {pages} {083107} (\bibinfo {year} {2019})},\
  \Eprint {http://arxiv.org/abs/https://doi.org/10.1063/1.5111934}
  {https://doi.org/10.1063/1.5111934} \BibitemShut {NoStop}%
\bibitem [{\citenamefont {Esirkepov}\ \emph {et~al.}(2014)\citenamefont
  {Esirkepov}, \citenamefont {Koga}, \citenamefont {Sunahara}, \citenamefont
  {Morita}, \citenamefont {Nishikino}, \citenamefont {Kageyama}, \citenamefont
  {Nagatomo}, \citenamefont {Nishihara}, \citenamefont {Sagisaka},
  \citenamefont {Kotaki}, \citenamefont {Nakamura}, \citenamefont {Fukuda},
  \citenamefont {Okada}, \citenamefont {Pirozhkov}, \citenamefont {Yogo},
  \citenamefont {Nishiuchi}, \citenamefont {Kiriyama}, \citenamefont {Kondo},
  \citenamefont {Kando},\ and\ \citenamefont {Bulanov}}]{esirkepov_preplasma}%
  \BibitemOpen
  \bibfield  {author} {\bibinfo {author} {\bibfnamefont {T.~Z.}\ \bibnamefont
  {Esirkepov}}, \bibinfo {author} {\bibfnamefont {J.~K.}\ \bibnamefont {Koga}},
  \bibinfo {author} {\bibfnamefont {A.}~\bibnamefont {Sunahara}}, \bibinfo
  {author} {\bibfnamefont {T.}~\bibnamefont {Morita}}, \bibinfo {author}
  {\bibfnamefont {M.}~\bibnamefont {Nishikino}}, \bibinfo {author}
  {\bibfnamefont {K.}~\bibnamefont {Kageyama}}, \bibinfo {author}
  {\bibfnamefont {H.}~\bibnamefont {Nagatomo}}, \bibinfo {author}
  {\bibfnamefont {K.}~\bibnamefont {Nishihara}}, \bibinfo {author}
  {\bibfnamefont {A.}~\bibnamefont {Sagisaka}}, \bibinfo {author}
  {\bibfnamefont {H.}~\bibnamefont {Kotaki}}, \bibinfo {author} {\bibfnamefont
  {T.}~\bibnamefont {Nakamura}}, \bibinfo {author} {\bibfnamefont
  {Y.}~\bibnamefont {Fukuda}}, \bibinfo {author} {\bibfnamefont
  {H.}~\bibnamefont {Okada}}, \bibinfo {author} {\bibfnamefont {A.~S.}\
  \bibnamefont {Pirozhkov}}, \bibinfo {author} {\bibfnamefont {A.}~\bibnamefont
  {Yogo}}, \bibinfo {author} {\bibfnamefont {M.}~\bibnamefont {Nishiuchi}},
  \bibinfo {author} {\bibfnamefont {H.}~\bibnamefont {Kiriyama}}, \bibinfo
  {author} {\bibfnamefont {K.}~\bibnamefont {Kondo}}, \bibinfo {author}
  {\bibfnamefont {M.}~\bibnamefont {Kando}}, \ and\ \bibinfo {author}
  {\bibfnamefont {S.~V.}\ \bibnamefont {Bulanov}},\ }\bibfield  {title}
  {\enquote {\bibinfo {title} {Prepulse and amplified spontaneous emission
  effects on the interaction of a petawatt class laser with thin solid
  targets},}\ }\href {\doibase https://doi.org/10.1016/j.nima.2014.01.056}
  {\bibfield  {journal} {\bibinfo  {journal} {Nuclear Instruments and Methods
  in Physics Research Section A: Accelerators, Spectrometers, Detectors and
  Associated Equipment}\ }\textbf {\bibinfo {volume} {745}},\ \bibinfo {pages}
  {150 -- 163} (\bibinfo {year} {2014})}\BibitemShut {NoStop}%
\bibitem [{\citenamefont {Hadjisolomou}\ \emph {et~al.}(2020)\citenamefont
  {Hadjisolomou}, \citenamefont {Tsygvintsev}, \citenamefont {Sasorov},
  \citenamefont {Gasilov}, \citenamefont {Korn},\ and\ \citenamefont
  {Bulanov}}]{preplasma_had}%
  \BibitemOpen
  \bibfield  {author} {\bibinfo {author} {\bibfnamefont {P.}~\bibnamefont
  {Hadjisolomou}}, \bibinfo {author} {\bibfnamefont {I.~P.}\ \bibnamefont
  {Tsygvintsev}}, \bibinfo {author} {\bibfnamefont {P.}~\bibnamefont
  {Sasorov}}, \bibinfo {author} {\bibfnamefont {V.}~\bibnamefont {Gasilov}},
  \bibinfo {author} {\bibfnamefont {G.}~\bibnamefont {Korn}}, \ and\ \bibinfo
  {author} {\bibfnamefont {S.~V.}\ \bibnamefont {Bulanov}},\ }\bibfield
  {title} {\enquote {\bibinfo {title} {Preplasma effects on laser ion
  generation from thin foil targets},}\ }\href {\doibase 10.1063/1.5124457}
  {\bibfield  {journal} {\bibinfo  {journal} {Physics of Plasmas}\ }\textbf
  {\bibinfo {volume} {27}},\ \bibinfo {pages} {013107} (\bibinfo {year}
  {2020})}\BibitemShut {NoStop}%
\bibitem [{\citenamefont {Paradkar}\ \emph {et~al.}(2011)\citenamefont
  {Paradkar}, \citenamefont {Wei}, \citenamefont {Yabuuchi}, \citenamefont
  {Stephens}, \citenamefont {Haines}, \citenamefont {Krasheninnikov},\ and\
  \citenamefont {Beg}}]{stoch_paradkar}%
  \BibitemOpen
  \bibfield  {author} {\bibinfo {author} {\bibfnamefont {B.~S.}\ \bibnamefont
  {Paradkar}}, \bibinfo {author} {\bibfnamefont {M.~S.}\ \bibnamefont {Wei}},
  \bibinfo {author} {\bibfnamefont {T.}~\bibnamefont {Yabuuchi}}, \bibinfo
  {author} {\bibfnamefont {R.~B.}\ \bibnamefont {Stephens}}, \bibinfo {author}
  {\bibfnamefont {M.~G.}\ \bibnamefont {Haines}}, \bibinfo {author}
  {\bibfnamefont {S.~I.}\ \bibnamefont {Krasheninnikov}}, \ and\ \bibinfo
  {author} {\bibfnamefont {F.~N.}\ \bibnamefont {Beg}},\ }\bibfield  {title}
  {\enquote {\bibinfo {title} {Numerical modeling of fast electron generation
  in the presence of preformed plasma in laser-matter interaction at
  relativistic intensities},}\ }\href {\doibase 10.1103/PhysRevE.83.046401}
  {\bibfield  {journal} {\bibinfo  {journal} {Phys. Rev. E}\ }\textbf {\bibinfo
  {volume} {83}},\ \bibinfo {pages} {046401} (\bibinfo {year}
  {2011})}\BibitemShut {NoStop}%
\bibitem [{\citenamefont {Sheng}\ \emph {et~al.}(2004)\citenamefont {Sheng},
  \citenamefont {Mima}, \citenamefont {Zhang},\ and\ \citenamefont
  {Meyer-Ter-Vehn}}]{stoch_sheng}%
  \BibitemOpen
  \bibfield  {author} {\bibinfo {author} {\bibfnamefont {Z.-M.}\ \bibnamefont
  {Sheng}}, \bibinfo {author} {\bibfnamefont {K.}~\bibnamefont {Mima}},
  \bibinfo {author} {\bibfnamefont {J.}~\bibnamefont {Zhang}}, \ and\ \bibinfo
  {author} {\bibfnamefont {J.}~\bibnamefont {Meyer-Ter-Vehn}},\ }\bibfield
  {title} {\enquote {\bibinfo {title} {Efficient acceleration of electrons with
  counterpropagating intense laser pulses in vacuum and underdense plasma},}\
  }\href {\doibase 10.1103/physreve.69.016407} {\bibfield  {journal} {\bibinfo
  {journal} {Physical review. E, Statistical, nonlinear, and soft matter
  physics}\ }\textbf {\bibinfo {volume} {69}},\ \bibinfo {pages} {016407}
  (\bibinfo {year} {2004})}\BibitemShut {NoStop}%
\bibitem [{\citenamefont {Sentoku}\ \emph {et~al.}(2002)\citenamefont
  {Sentoku}, \citenamefont {Bychenkov}, \citenamefont {Flippo}, \citenamefont
  {Maksimchuk}, \citenamefont {Mima}, \citenamefont {Mourou}, \citenamefont
  {Sheng},\ and\ \citenamefont {Umstadter}}]{sentoku_stoch}%
  \BibitemOpen
  \bibfield  {author} {\bibinfo {author} {\bibfnamefont {Y.}~\bibnamefont
  {Sentoku}}, \bibinfo {author} {\bibfnamefont {V.}~\bibnamefont {Bychenkov}},
  \bibinfo {author} {\bibfnamefont {K.}~\bibnamefont {Flippo}}, \bibinfo
  {author} {\bibfnamefont {A.}~\bibnamefont {Maksimchuk}}, \bibinfo {author}
  {\bibfnamefont {K.}~\bibnamefont {Mima}}, \bibinfo {author} {\bibfnamefont
  {G.}~\bibnamefont {Mourou}}, \bibinfo {author} {\bibfnamefont
  {Z.}~\bibnamefont {Sheng}}, \ and\ \bibinfo {author} {\bibfnamefont
  {D.}~\bibnamefont {Umstadter}},\ }\bibfield  {title} {\enquote {\bibinfo
  {title} {High-energy ion generation in interaction of short laser pulse with
  high-density plasma},}\ }\href {\doibase 10.1007/s003400200796} {\bibfield
  {journal} {\bibinfo  {journal} {Applied Physics B}\ }\textbf {\bibinfo
  {volume} {74}},\ \bibinfo {pages} {207--} (\bibinfo {year}
  {2002})}\BibitemShut {NoStop}%
\bibitem [{\citenamefont {Chopineau}\ \emph {et~al.}(2019)\citenamefont
  {Chopineau}, \citenamefont {Leblanc}, \citenamefont {Blaclard}, \citenamefont
  {Denoeud}, \citenamefont {Th\'evenet}, \citenamefont {Vay}, \citenamefont
  {Bonnaud}, \citenamefont {Martin}, \citenamefont {Vincenti},\ and\
  \citenamefont {Qu\'er\'e}}]{chopineau_stoch}%
  \BibitemOpen
  \bibfield  {author} {\bibinfo {author} {\bibfnamefont {L.}~\bibnamefont
  {Chopineau}}, \bibinfo {author} {\bibfnamefont {A.}~\bibnamefont {Leblanc}},
  \bibinfo {author} {\bibfnamefont {G.}~\bibnamefont {Blaclard}}, \bibinfo
  {author} {\bibfnamefont {A.}~\bibnamefont {Denoeud}}, \bibinfo {author}
  {\bibfnamefont {M.}~\bibnamefont {Th\'evenet}}, \bibinfo {author}
  {\bibfnamefont {J.-L.}\ \bibnamefont {Vay}}, \bibinfo {author} {\bibfnamefont
  {G.}~\bibnamefont {Bonnaud}}, \bibinfo {author} {\bibfnamefont
  {P.}~\bibnamefont {Martin}}, \bibinfo {author} {\bibfnamefont
  {H.}~\bibnamefont {Vincenti}}, \ and\ \bibinfo {author} {\bibfnamefont
  {F.}~\bibnamefont {Qu\'er\'e}},\ }\bibfield  {title} {\enquote {\bibinfo
  {title} {Identification of coupling mechanisms between ultraintense laser
  light and dense plasmas},}\ }\href {\doibase 10.1103/PhysRevX.9.011050}
  {\bibfield  {journal} {\bibinfo  {journal} {Phys. Rev. X}\ }\textbf {\bibinfo
  {volume} {9}},\ \bibinfo {pages} {011050} (\bibinfo {year}
  {2019})}\BibitemShut {NoStop}%
\bibitem [{\citenamefont {Pukhov}, \citenamefont {Sheng},\ and\ \citenamefont
  {Meyer-ter Vehn}(1999)}]{dla}%
  \BibitemOpen
  \bibfield  {author} {\bibinfo {author} {\bibfnamefont {A.}~\bibnamefont
  {Pukhov}}, \bibinfo {author} {\bibfnamefont {Z.-M.}\ \bibnamefont {Sheng}}, \
  and\ \bibinfo {author} {\bibfnamefont {J.}~\bibnamefont {Meyer-ter Vehn}},\
  }\bibfield  {title} {\enquote {\bibinfo {title} {Particle acceleration in
  relativistic laser channels},}\ }\href {\doibase 10.1063/1.873242} {\bibfield
   {journal} {\bibinfo  {journal} {Physics of Plasmas}\ }\textbf {\bibinfo
  {volume} {6}},\ \bibinfo {pages} {2847--2854} (\bibinfo {year} {1999})},\
  \Eprint {http://arxiv.org/abs/https://doi.org/10.1063/1.873242}
  {https://doi.org/10.1063/1.873242} \BibitemShut {NoStop}%
\bibitem [{\citenamefont {Krygier}, \citenamefont {Schumacher},\ and\
  \citenamefont {Freeman}(2014)}]{lida}%
  \BibitemOpen
  \bibfield  {author} {\bibinfo {author} {\bibfnamefont {A.~G.}\ \bibnamefont
  {Krygier}}, \bibinfo {author} {\bibfnamefont {D.~W.}\ \bibnamefont
  {Schumacher}}, \ and\ \bibinfo {author} {\bibfnamefont {R.~R.}\ \bibnamefont
  {Freeman}},\ }\bibfield  {title} {\enquote {\bibinfo {title} {On the origin
  of super-hot electrons from intense laser interactions with solid targets
  having moderate scale length preformed plasmas},}\ }\href {\doibase
  10.1063/1.4866587} {\bibfield  {journal} {\bibinfo  {journal} {Physics of
  Plasmas}\ }\textbf {\bibinfo {volume} {21}},\ \bibinfo {pages} {023112}
  (\bibinfo {year} {2014})},\ \Eprint
  {http://arxiv.org/abs/https://doi.org/10.1063/1.4866587}
  {https://doi.org/10.1063/1.4866587} \BibitemShut {NoStop}%
\bibitem [{\citenamefont {Derouillat}\ \emph {et~al.}(2018)\citenamefont
  {Derouillat}, \citenamefont {Beck}, \citenamefont {Pérez}, \citenamefont
  {Vinci}, \citenamefont {Chiaramello}, \citenamefont {Grassi}, \citenamefont
  {Flé}, \citenamefont {Bouchard}, \citenamefont {Plotnikov}, \citenamefont
  {Aunai}, \citenamefont {Dargent}, \citenamefont {Riconda},\ and\
  \citenamefont {Grech}}]{smilei}%
  \BibitemOpen
  \bibfield  {author} {\bibinfo {author} {\bibfnamefont {J.}~\bibnamefont
  {Derouillat}}, \bibinfo {author} {\bibfnamefont {A.}~\bibnamefont {Beck}},
  \bibinfo {author} {\bibfnamefont {F.}~\bibnamefont {Pérez}}, \bibinfo
  {author} {\bibfnamefont {T.}~\bibnamefont {Vinci}}, \bibinfo {author}
  {\bibfnamefont {M.}~\bibnamefont {Chiaramello}}, \bibinfo {author}
  {\bibfnamefont {A.}~\bibnamefont {Grassi}}, \bibinfo {author} {\bibfnamefont
  {M.}~\bibnamefont {Flé}}, \bibinfo {author} {\bibfnamefont {G.}~\bibnamefont
  {Bouchard}}, \bibinfo {author} {\bibfnamefont {I.}~\bibnamefont {Plotnikov}},
  \bibinfo {author} {\bibfnamefont {N.}~\bibnamefont {Aunai}}, \bibinfo
  {author} {\bibfnamefont {J.}~\bibnamefont {Dargent}}, \bibinfo {author}
  {\bibfnamefont {C.}~\bibnamefont {Riconda}}, \ and\ \bibinfo {author}
  {\bibfnamefont {M.}~\bibnamefont {Grech}},\ }\bibfield  {title} {\enquote
  {\bibinfo {title} {Smilei : A collaborative, open-source, multi-purpose
  particle-in-cell code for plasma simulation},}\ }\href {\doibase
  https://doi.org/10.1016/j.cpc.2017.09.024} {\bibfield  {journal} {\bibinfo
  {journal} {Computer Physics Communications}\ }\textbf {\bibinfo {volume}
  {222}},\ \bibinfo {pages} {351 -- 373} (\bibinfo {year} {2018})}\BibitemShut
  {NoStop}%
\bibitem [{\citenamefont {Liseykina}, \citenamefont {Mulser},\ and\
  \citenamefont {Murakami}(2015)}]{liseykina_dist}%
  \BibitemOpen
  \bibfield  {author} {\bibinfo {author} {\bibfnamefont {T.}~\bibnamefont
  {Liseykina}}, \bibinfo {author} {\bibfnamefont {P.}~\bibnamefont {Mulser}}, \
  and\ \bibinfo {author} {\bibfnamefont {M.}~\bibnamefont {Murakami}},\
  }\bibfield  {title} {\enquote {\bibinfo {title} {Collisionless absorption,
  hot electron generation, and energy scaling in intense laser-target
  interaction},}\ }\href {\doibase 10.1063/1.4914837} {\bibfield  {journal}
  {\bibinfo  {journal} {Physics of Plasmas}\ }\textbf {\bibinfo {volume}
  {22}},\ \bibinfo {pages} {033302} (\bibinfo {year} {2015})},\ \Eprint
  {http://arxiv.org/abs/https://doi.org/10.1063/1.4914837}
  {https://doi.org/10.1063/1.4914837} \BibitemShut {NoStop}%
\bibitem [{\citenamefont {Tikhonchuk}\ \emph {et~al.}(2005)\citenamefont
  {Tikhonchuk}, \citenamefont {Andreev}, \citenamefont {Bochkarev},\ and\
  \citenamefont {Bychenkov}}]{tikhonchuk_ionacc}%
  \BibitemOpen
  \bibfield  {author} {\bibinfo {author} {\bibfnamefont {V.~T.}\ \bibnamefont
  {Tikhonchuk}}, \bibinfo {author} {\bibfnamefont {A.~A.}\ \bibnamefont
  {Andreev}}, \bibinfo {author} {\bibfnamefont {S.~G.}\ \bibnamefont
  {Bochkarev}}, \ and\ \bibinfo {author} {\bibfnamefont {V.~Y.}\ \bibnamefont
  {Bychenkov}},\ }\bibfield  {title} {\enquote {\bibinfo {title} {Ion
  acceleration in short-laser-pulse interaction with solid foils},}\ }\href
  {\doibase 10.1088/0741-3335/47/12b/s69} {\bibfield  {journal} {\bibinfo
  {journal} {Plasma Physics and Controlled Fusion}\ }\textbf {\bibinfo {volume}
  {47}},\ \bibinfo {pages} {B869--B877} (\bibinfo {year} {2005})}\BibitemShut
  {NoStop}%
\bibitem [{\citenamefont {Cai}\ \emph {et~al.}(2006)\citenamefont {Cai},
  \citenamefont {Yu}, \citenamefont {Zhu},\ and\ \citenamefont
  {Zheng}}]{cai_jxb}%
  \BibitemOpen
  \bibfield  {author} {\bibinfo {author} {\bibfnamefont {H.-B.}\ \bibnamefont
  {Cai}}, \bibinfo {author} {\bibfnamefont {W.}~\bibnamefont {Yu}}, \bibinfo
  {author} {\bibfnamefont {S.-P.}\ \bibnamefont {Zhu}}, \ and\ \bibinfo
  {author} {\bibfnamefont {C.-Y.}\ \bibnamefont {Zheng}},\ }\bibfield  {title}
  {\enquote {\bibinfo {title} {Short-pulse laser absorption via jx$\rm{B}$
  heating in ultrahigh intensity laser plasma interaction},}\ }\href {\doibase
  10.1063/1.2372463} {\bibfield  {journal} {\bibinfo  {journal} {Physics of
  Plasmas}\ }\textbf {\bibinfo {volume} {13}},\ \bibinfo {pages}
  {113105--113105} (\bibinfo {year} {2006})}\BibitemShut {NoStop}%
\bibitem [{\citenamefont {Wagner}\ \emph {et~al.}(2016)\citenamefont {Wagner},
  \citenamefont {Brabetz}, \citenamefont {Deppert}, \citenamefont {Roth},
  \citenamefont {Stoehlker}, \citenamefont {Tauschwitz}, \citenamefont
  {Tebartz}, \citenamefont {Zielbauer},\ and\ \citenamefont
  {Bagnoud}}]{Wagner2016}%
  \BibitemOpen
  \bibfield  {author} {\bibinfo {author} {\bibfnamefont {F.}~\bibnamefont
  {Wagner}}, \bibinfo {author} {\bibfnamefont {C.}~\bibnamefont {Brabetz}},
  \bibinfo {author} {\bibfnamefont {O.}~\bibnamefont {Deppert}}, \bibinfo
  {author} {\bibfnamefont {M.}~\bibnamefont {Roth}}, \bibinfo {author}
  {\bibfnamefont {T.}~\bibnamefont {Stoehlker}}, \bibinfo {author}
  {\bibfnamefont {A.}~\bibnamefont {Tauschwitz}}, \bibinfo {author}
  {\bibfnamefont {A.}~\bibnamefont {Tebartz}}, \bibinfo {author} {\bibfnamefont
  {B.}~\bibnamefont {Zielbauer}}, \ and\ \bibinfo {author} {\bibfnamefont
  {V.}~\bibnamefont {Bagnoud}},\ }\bibfield  {title} {\enquote {\bibinfo
  {title} {Accelerating ions with high-energy short laser pulses from
  submicrometer thick targets},}\ }\href {\doibase 10.1017/hpl.2016.44}
  {\bibfield  {journal} {\bibinfo  {journal} {High Power Laser Sci. Eng.}\
  }\textbf {\bibinfo {volume} {4}},\ \bibinfo {pages} {e45} (\bibinfo {year}
  {2016})}\BibitemShut {NoStop}%
\bibitem [{\citenamefont {Klimo}\ \emph {et~al.}(2008)\citenamefont {Klimo},
  \citenamefont {Psikal}, \citenamefont {Limpouch},\ and\ \citenamefont
  {Tikhonchuk}}]{klimo_hb}%
  \BibitemOpen
  \bibfield  {author} {\bibinfo {author} {\bibfnamefont {O.}~\bibnamefont
  {Klimo}}, \bibinfo {author} {\bibfnamefont {J.}~\bibnamefont {Psikal}},
  \bibinfo {author} {\bibfnamefont {J.}~\bibnamefont {Limpouch}}, \ and\
  \bibinfo {author} {\bibfnamefont {V.~T.}\ \bibnamefont {Tikhonchuk}},\
  }\bibfield  {title} {\enquote {\bibinfo {title} {Monoenergetic ion beams from
  ultrathin foils irradiated by ultrahigh-contrast circularly polarized laser
  pulses},}\ }\href {\doibase 10.1103/PhysRevSTAB.11.031301} {\bibfield
  {journal} {\bibinfo  {journal} {Phys. Rev. ST Accel. Beams}\ }\textbf
  {\bibinfo {volume} {11}},\ \bibinfo {pages} {031301} (\bibinfo {year}
  {2008})}\BibitemShut {NoStop}%
\bibitem [{\citenamefont {Robinson}\ \emph {et~al.}(2009)\citenamefont
  {Robinson}, \citenamefont {Gibbon}, \citenamefont {Zepf}, \citenamefont
  {Kar}, \citenamefont {Evans},\ and\ \citenamefont {Bellei}}]{Robinson2009}%
  \BibitemOpen
  \bibfield  {author} {\bibinfo {author} {\bibfnamefont {A.~P.~L.}\
  \bibnamefont {Robinson}}, \bibinfo {author} {\bibfnamefont {P.}~\bibnamefont
  {Gibbon}}, \bibinfo {author} {\bibfnamefont {M.}~\bibnamefont {Zepf}},
  \bibinfo {author} {\bibfnamefont {S.}~\bibnamefont {Kar}}, \bibinfo {author}
  {\bibfnamefont {R.~G.}\ \bibnamefont {Evans}}, \ and\ \bibinfo {author}
  {\bibfnamefont {C.}~\bibnamefont {Bellei}},\ }\bibfield  {title} {\enquote
  {\bibinfo {title} {Relativistically correct hole-boring and ion acceleration
  by circularly polarized laser pulses},}\ }\href {\doibase
  10.1088/0741-3335/51/2/024004} {\bibfield  {journal} {\bibinfo  {journal}
  {Plasma Phys. Control. Fusion}\ }\textbf {\bibinfo {volume} {51}},\ \bibinfo
  {pages} {024004} (\bibinfo {year} {2009})}\BibitemShut {NoStop}%
\bibitem [{\citenamefont {Cattani}\ \emph {et~al.}(2000)\citenamefont
  {Cattani}, \citenamefont {Kim}, \citenamefont {Anderson},\ and\ \citenamefont
  {Lisak}}]{Cattani2000}%
  \BibitemOpen
  \bibfield  {author} {\bibinfo {author} {\bibfnamefont {F.}~\bibnamefont
  {Cattani}}, \bibinfo {author} {\bibfnamefont {A.}~\bibnamefont {Kim}},
  \bibinfo {author} {\bibfnamefont {D.}~\bibnamefont {Anderson}}, \ and\
  \bibinfo {author} {\bibfnamefont {M.}~\bibnamefont {Lisak}},\ }\bibfield
  {title} {\enquote {\bibinfo {title} {Threshold of induced transparency in the
  relativistic interaction of an electromagnetic wave with overdense
  plasmas},}\ }\href {\doibase 10.1103/PhysRevE.62.1234} {\bibfield  {journal}
  {\bibinfo  {journal} {Phys. Rev. E}\ }\textbf {\bibinfo {volume} {62}},\
  \bibinfo {pages} {1234--1237} (\bibinfo {year} {2000})}\BibitemShut {NoStop}%
\bibitem [{\citenamefont {{Lichtenberg A. J. and Lieberman M.
  A.}}(1983)}]{lieberman_stoch}%
  \BibitemOpen
  \bibfield  {author} {\bibinfo {author} {\bibnamefont {{Lichtenberg A. J. and
  Lieberman M. A.}}},\ }\href {\doibase 10.1007/978-1-4757-4257-2} {\emph
  {\bibinfo {title} {Regular nad stochastic motion}}}\ (\bibinfo  {publisher}
  {Springer Science + Business Media New York},\ \bibinfo {year}
  {1983})\BibitemShut {NoStop}%
\bibitem [{\citenamefont {Stark}\ \emph {et~al.}(2017)\citenamefont {Stark},
  \citenamefont {Yin}, \citenamefont {Albright},\ and\ \citenamefont
  {Guo}}]{Stark2017}%
  \BibitemOpen
  \bibfield  {author} {\bibinfo {author} {\bibfnamefont {D.~J.}\ \bibnamefont
  {Stark}}, \bibinfo {author} {\bibfnamefont {L.}~\bibnamefont {Yin}}, \bibinfo
  {author} {\bibfnamefont {B.~J.}\ \bibnamefont {Albright}}, \ and\ \bibinfo
  {author} {\bibfnamefont {F.}~\bibnamefont {Guo}},\ }\bibfield  {title}
  {\enquote {\bibinfo {title} {Effects of dimensionality on kinetic simulations
  of laser-ion acceleration in the transparency regime},}\ }\href {\doibase
  10.1063/1.4982741} {\bibfield  {journal} {\bibinfo  {journal} {Phys.
  Plasmas}\ }\textbf {\bibinfo {volume} {24}},\ \bibinfo {pages} {053103}
  (\bibinfo {year} {2017})}\BibitemShut {NoStop}%
\bibitem [{\citenamefont {Kaplan}\ and\ \citenamefont
  {Pokrovsky}(2005)}]{kaplan_sw}%
  \BibitemOpen
  \bibfield  {author} {\bibinfo {author} {\bibfnamefont {A.}~\bibnamefont
  {Kaplan}}\ and\ \bibinfo {author} {\bibfnamefont {A.}~\bibnamefont
  {Pokrovsky}},\ }\bibfield  {title} {\enquote {\bibinfo {title} {Fully
  relativistic theory of the ponderomotive force in an ultraintense standing
  wave},}\ }\href {\doibase 10.1103/PhysRevLett.95.053601} {\bibfield
  {journal} {\bibinfo  {journal} {Physical review letters}\ }\textbf {\bibinfo
  {volume} {95}},\ \bibinfo {pages} {053601} (\bibinfo {year}
  {2005})}\BibitemShut {NoStop}%
\bibitem [{\citenamefont {Bauer}, \citenamefont {Mulser},\ and\ \citenamefont
  {Steeb}(1995)}]{bauer_pond}%
  \BibitemOpen
  \bibfield  {author} {\bibinfo {author} {\bibfnamefont {D.}~\bibnamefont
  {Bauer}}, \bibinfo {author} {\bibfnamefont {P.}~\bibnamefont {Mulser}}, \
  and\ \bibinfo {author} {\bibfnamefont {W.~H.}\ \bibnamefont {Steeb}},\
  }\bibfield  {title} {\enquote {\bibinfo {title} {Relativistic ponderomotive
  force, uphill acceleration, and transition to chaos},}\ }\href {\doibase
  10.1103/PhysRevLett.75.4622} {\bibfield  {journal} {\bibinfo  {journal}
  {Phys. Rev. Lett.}\ }\textbf {\bibinfo {volume} {75}},\ \bibinfo {pages}
  {4622--4625} (\bibinfo {year} {1995})}\BibitemShut {NoStop}%
\bibitem [{\citenamefont {Lehmann}\ and\ \citenamefont
  {Spatschek}(2012)}]{lehmann_attractors}%
  \BibitemOpen
  \bibfield  {author} {\bibinfo {author} {\bibfnamefont {G.}~\bibnamefont
  {Lehmann}}\ and\ \bibinfo {author} {\bibfnamefont {K.~H.}\ \bibnamefont
  {Spatschek}},\ }\bibfield  {title} {\enquote {\bibinfo {title} {Phase-space
  contraction and attractors for ultrarelativistic electrons},}\ }\href
  {\doibase 10.1103/PhysRevE.85.056412} {\bibfield  {journal} {\bibinfo
  {journal} {Phys. Rev. E}\ }\textbf {\bibinfo {volume} {85}},\ \bibinfo
  {pages} {056412} (\bibinfo {year} {2012})}\BibitemShut {NoStop}%
\bibitem [{\citenamefont {Sheng}\ \emph {et~al.}(2002)\citenamefont {Sheng},
  \citenamefont {Mima}, \citenamefont {Sentoku}, \citenamefont
  {Jovanovi\ifmmode~\acute{c}\else \'{c}\fi{}}, \citenamefont {Taguchi},
  \citenamefont {Zhang},\ and\ \citenamefont {Meyer-ter Vehn}}]{sheng_stoch}%
  \BibitemOpen
  \bibfield  {author} {\bibinfo {author} {\bibfnamefont {Z.-M.}\ \bibnamefont
  {Sheng}}, \bibinfo {author} {\bibfnamefont {K.}~\bibnamefont {Mima}},
  \bibinfo {author} {\bibfnamefont {Y.}~\bibnamefont {Sentoku}}, \bibinfo
  {author} {\bibfnamefont {M.~S.}\ \bibnamefont
  {Jovanovi\ifmmode~\acute{c}\else \'{c}\fi{}}}, \bibinfo {author}
  {\bibfnamefont {T.}~\bibnamefont {Taguchi}}, \bibinfo {author} {\bibfnamefont
  {J.}~\bibnamefont {Zhang}}, \ and\ \bibinfo {author} {\bibfnamefont
  {J.}~\bibnamefont {Meyer-ter Vehn}},\ }\bibfield  {title} {\enquote {\bibinfo
  {title} {Stochastic heating and acceleration of electrons in colliding laser
  fields in plasma},}\ }\href {\doibase 10.1103/PhysRevLett.88.055004}
  {\bibfield  {journal} {\bibinfo  {journal} {Phys. Rev. Lett.}\ }\textbf
  {\bibinfo {volume} {88}},\ \bibinfo {pages} {055004} (\bibinfo {year}
  {2002})}\BibitemShut {NoStop}%
\bibitem [{\citenamefont {Khudik}\ \emph {et~al.}(2016)\citenamefont {Khudik},
  \citenamefont {Arefiev}, \citenamefont {Zhang},\ and\ \citenamefont
  {Shvets}}]{khudik_channel}%
  \BibitemOpen
  \bibfield  {author} {\bibinfo {author} {\bibfnamefont {V.}~\bibnamefont
  {Khudik}}, \bibinfo {author} {\bibfnamefont {A.}~\bibnamefont {Arefiev}},
  \bibinfo {author} {\bibfnamefont {X.}~\bibnamefont {Zhang}}, \ and\ \bibinfo
  {author} {\bibfnamefont {G.}~\bibnamefont {Shvets}},\ }\bibfield  {title}
  {\enquote {\bibinfo {title} {Universal scalings for laser acceleration of
  electrons in ion channels},}\ }\href {\doibase 10.1063/1.4964901} {\bibfield
  {journal} {\bibinfo  {journal} {Physics of Plasmas}\ }\textbf {\bibinfo
  {volume} {23}},\ \bibinfo {pages} {103108} (\bibinfo {year}
  {2016})}\BibitemShut {NoStop}%
\bibitem [{\citenamefont {Vranic}, \citenamefont {Fonseca},\ and\ \citenamefont
  {Silva}(2018)}]{vranic_channel}%
  \BibitemOpen
  \bibfield  {author} {\bibinfo {author} {\bibfnamefont {M.}~\bibnamefont
  {Vranic}}, \bibinfo {author} {\bibfnamefont {R.~A.}\ \bibnamefont {Fonseca}},
  \ and\ \bibinfo {author} {\bibfnamefont {L.~O.}\ \bibnamefont {Silva}},\
  }\bibfield  {title} {\enquote {\bibinfo {title} {Extremely intense
  laser-based electron acceleration in a plasma channel},}\ }\href {\doibase
  10.1088/1361-6587/aaa36c} {\bibfield  {journal} {\bibinfo  {journal} {Plasma
  Physics and Controlled Fusion}\ }\textbf {\bibinfo {volume} {60}},\ \bibinfo
  {pages} {034002} (\bibinfo {year} {2018})}\BibitemShut {NoStop}%
\end{thebibliography}%
	
\end{document}